\documentclass[%
preprint,
nofootinbib,
 amsmath,amssymb,
 aps,
]{revtex4-1}

\usepackage{dcolumn}

\allowdisplaybreaks

\usepackage[dvipdfmx]{graphicx}
\usepackage{latexsym}
\usepackage{amsfonts}
\usepackage{amssymb}
\usepackage{amsmath}
\usepackage{bm}
\usepackage{mathrsfs} 

\newcommand{\bx}{{\bm{x}}}

\newcommand{\bq}{{\bm{q}}}
\newcommand{\integ}{\int\!\!}

\newcommand{\Pex}{{\Pi_{ {\rm ex} }}}
\newcommand{\Pexr}{ \hat \Pi_{\rm ex}  }

\newcommand{\chiv}{ \chi_0^{\rm vac} }
\newcommand{\chivr}{ \hat \chi_0^{\rm vac} }

\newcommand{\R}{\mathcal R}

\newcommand{\bB}{{\bm{B}}}
\newcommand{\bE}{{\bm{E}}}

\newcommand{\Lag}{ {\mathscr{L}} }

\newcommand{\bt}{{\beta\tau}}
\newcommand{\Br}{B_{\rm r}}

\newcommand{\rp}{{r_\parallel^2}}
\newcommand{\rt}{{r_\perp^2}}
\newcommand{\qp}{{q_\parallel^2}}
\newcommand{\qt}{{q_\perp^2}}

\newcommand{\sgn}{{\rm sgn}}

\newcommand{\idx}{ { {}_{\ell n} } }

\newcommand{\ome}{\tilde \omega}

\newcommand{\eperp}{\epsilon_\perp}
\newcommand{\epara}{\epsilon_\parallel}

\newcommand{\C}{ C_\ell^n }

\def\simge{\mathrel{%
   \rlap{\raise 0.511ex \hbox{$>$}}{\lower 0.511ex \hbox{$\sim$}}}}
\def\simle{\mathrel{
   \rlap{\raise 0.511ex \hbox{$<$}}{\lower 0.511ex \hbox{$\sim$}}}}
\def\bigs{\mathrel{
   \rlap{\raise 0.531ex \hbox{$>$}}{\lower 0.531ex \hbox{$<$}}}}

\usepackage[normalem]{ulem}  
\usepackage[dvips]{color} 

\renewcommand\sout{\bgroup \color{red} \ULdepth=-.5ex \ULset}

\usepackage{slashed}

\begin{document} 




\author{Koichi Hattori}\email{\tt khattori@yonsei.ac.kr}

\affiliation{Theory Center, IPNS, 
High energy accelerator research organization (KEK), 
1-1 Oho, Tsukuba, Ibaraki 305-0801, Japan}
\affiliation{Institute of Physics and Applied Physics, 
Yonsei University, Seoul 120-749, Korea \footnote{current address}}

\author{Kazunori Itakura }\email{\tt kazunori.itakura@kek.jp}
\affiliation{Theory Center, IPNS, 
High energy accelerator research organization (KEK), 
1-1 Oho, Tsukuba, Ibaraki 305-0801, Japan}
\affiliation{Department of Particle and Nuclear Studies, 
Graduate University for Advanced Studies (SOKENDAI), 
1-1 Oho, Tsukuba, Ibaraki 305-0801, Japan}


\title{Vacuum birefringence in strong magnetic fields: \\
(I) Photon polarization tensor with all the Landau levels}



\begin{abstract}
Photons propagating in strong magnetic fields are subject to a phenomenon 
called the ``vacuum birefringence" where refractive indices of two physical 
modes both deviate from unity and are different from each other. We compute 
the vacuum polarization tensor of a photon in a static and homogeneous 
magnetic field by utilizing Schwinger's proper-time method, and obtain a 
series representation as a result of double integrals {\it analytically} 
performed with respect to proper-time variables. The outcome is expressed 
in terms of an infinite sum of known functions which is plausibly interpreted 
as summation over all the Landau levels of fermions. Each 
contribution from infinitely many Landau levels yields a kinematical 
condition above which the contribution has an imaginary part. This 
indicates decay of a sufficiently energetic photon into a 
fermion-antifermion pair with corresponding Landau level indices. 
Since we do not resort to any approximation, 
our result is applicable to the calculation of refractive indices 
in the whole kinematical region of a photon momentum and in 
any magnitude of the external magnetic field. 

\end{abstract}


\preprint{KEK-TH-1567 }



\maketitle


\section{Introduction}
\label{sec:intro}


It has been long studied that structure of the quantum 
vacuum in QED would be modified in the presence of externally 
applied strong electromagnetic fields \cite{HE,Dun}, and 
modification of the vacuum could entail novel phenomena such 
as vacuum birefringence of a photon, photon decay into an 
electron-positron pair, and photon splitting. 
It is natural to expect these effects to occur because the 
vacuum in QED is filled with electrons in the Dirac sea, 
and they react as `media' like in ordinary substances
in response to external fields \cite{Weisskopf}. 
Whereas any such effect has not been established 
in experiments, recent years have witnessed an increasing 
interest in possibilities that extremely strong electromagnetic 
fields would be realized in several different situations 
(see, for example, Ref.~\cite{PIF} for a wide range of 
physics related to strong fields). 
Primary examples are relativistic (non-central) heavy-ion 
collisions \cite{KMW,Sko,DH,Itakura_PIF} and 
strongly-magnetized compact stars such as magnetars \cite{TD}, 
both of which are thought to accompany 
electromagnetic fields far stronger than 
so-called the ``critical field" for electrons,\footnote{
Since the magnetic axis of a magnetar is generally  
not the same as the rotation axis (thus, a pulsar), the magnetic 
field shows strong time dependence and will induce an electric field 
of the similar order of strength.} 
$B_c=E_c\equiv m_e^2/|e|$. Many observable effects are proposed 
in relation to magnetars \cite{MagnetarReview} and also to 
heavy-ion collisions \cite{Tu,Itakura_PIF,IHhbt}. Besides, intensity 
of high-field laser has been 
rapidly increasing and is now about to reach the critical field strength. 
Many authors have been addressing theoretical aspects of 
the possible vacuum physics that could be studied 
by the ultra-high-intensity laser (see for example, 
Ref.~\cite{TopicalReviewLASER}). 
Thus, in near future, we would
be able to study the vacuum physics in strong fields 
in somewhat controllable environments. With these 
current situations in mind, we should further pursue
towards deeper 
understanding of the vacuum physics in strong fields.

In the present paper, we provide a theoretical framework necessary 
for the description of 
vacuum birefringence and photon decay in strong magnetic
fields which are typical and important examples of the vacuum physics in 
strong fields. We give an {\it analytic} expression for the 
vacuum polarization tensor of a photon within one loop of a `dressed' 
fermion that includes all-order interaction with the external field at 
tree level. Notice that the strong fields compensate 
the smallness of the coupling constant, 
making a particular kind of higher-order diagrams enhanced. 
Moreover, when the fields are stronger than the critical ones, 
na\"ive perturbative expansion breaks down and 
thus one has to sum up all-order diagrams that are enhanced by the 
external fields. The dressed fermion propagator which appears in 
the one-loop diagram is obtained after such resummation with respect 
to the external fields.
As a result of resummation of higher-order diagrams, 
observables acquire nonlinear dependences on the external field. 
Since interaction with the external field is encoded as a linear 
coupling at the Lagrangian level, 
strong-field physics describing this kind of nonlinear effect 
is called the ``nonlinear QED".

In spite of a simple structure of the one-loop diagram, complete
description of the vacuum polarization tensor in the whole kinematical 
region has not been available so far. While the vacuum 
polarization tensor at the one-loop level is given in an integral 
representation with respect to two proper-time variables 
\cite{Tsa,Urr78,Adl1,TE,MS76,DR,DG}, 
its integrand is composed of rapidly oscillating exponential factors 
which have prevented previous studies from precise analytic and even 
numerical understanding of the phenomena. 
Because of this difficulty, 
our understanding of the vacuum birefringence was limited to a 
restricted phase space of the photon momentum \cite{Adl1,TE,KY} 
or to a strong-field case where lowest Landau level approximation
is applicable
\cite{MS76}.
We show however that the double proper-time integral 
can be analytically performed owing to 
a double infinite series expansions of the integrand. 
Since our calculation does not resort to any 
approximation, the result is applicable to any value of 
a photon momentum and any magnitude of the external field. 
Namely, our result can also treat, as well as the vacuum birefringence, the photon decay into 
a fermion-antifermion pair which occurs when the photon energy is 
sufficiently large.

This paper is organized as follows. First, in Sec.~\ref{sec:vp}, 
we present the integral representation of the vacuum polarization tensor 
of a photon in the external magnetic field. Then, in Sec.~\ref{sec:optics},
we explain how the vacuum birefringence appears in strong 
magnetic fields. Analytic evaluation of the integral is discussed in 
detail in Sec.~\ref{sec:SeriesExpansion}. In this section, we also provide 
physical interpretation of the double infinite series expansion, and 
then carefully investigate singularity structure of the vacuum polarization 
tensor. Kinematical condition of the photon decay also appears here. 
Summary and prospects are given in the last section 
where we mention another important step to obtain the refractive indices. 
In Appendices, we explain supplemental materials such as details of the 
proper-time method, renormalization issues, and so on. 
In principle, we are able to calculate the refractive indices by
using the polarization tensor obtained in the present paper. 
Taking care of the procedure mentioned in the last section, 
we will show the results of refractive indices in the next 
paper \cite{HI2}.


\section{Vacuum polarization tensor in external fields}
\label{sec:vp}


In this section, we provide theoretical framework for 
the vacuum polarization tensor of a propagating photon in a strong 
external field, 
which is necessary for the calculation of the refractive indices. 
In particular, we discuss the case of a strong {\it magnetic} field, but the 
techniques developed in this section should be equally applied to a 
case with a strong {\it electric} field, which will be reported separately.

\begin{figure}
\begin{minipage}{0.5\hsize} 
	\begin{center}\hspace*{-5mm}
		\includegraphics[width=0.999\hsize]{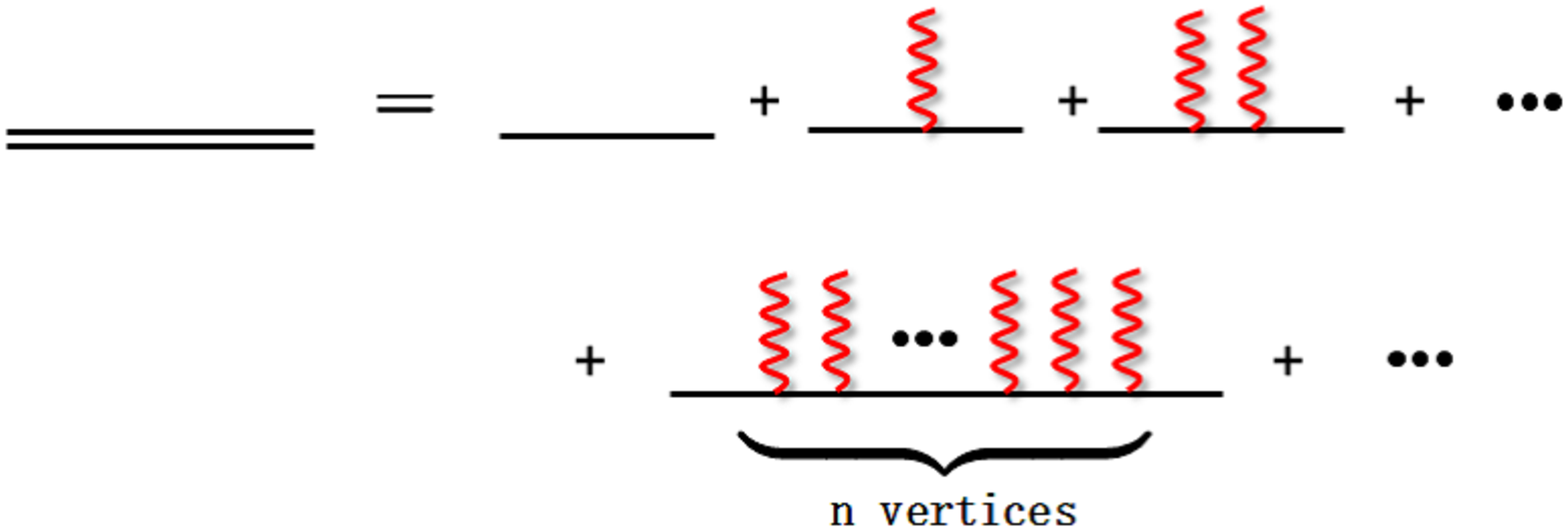}
	\end{center}
\vspace{-0.8cm}
\caption{Dressed fermion propagator (a double line) includes all 
the tree-level interactions with a strong external field (wavy lines).}
\label{fig:DressedProp}
\end{minipage}
\quad\ 
\begin{minipage}{0.36\hsize}
	\begin{center}\vspace*{0.65cm}
		\includegraphics[width=0.9\hsize]{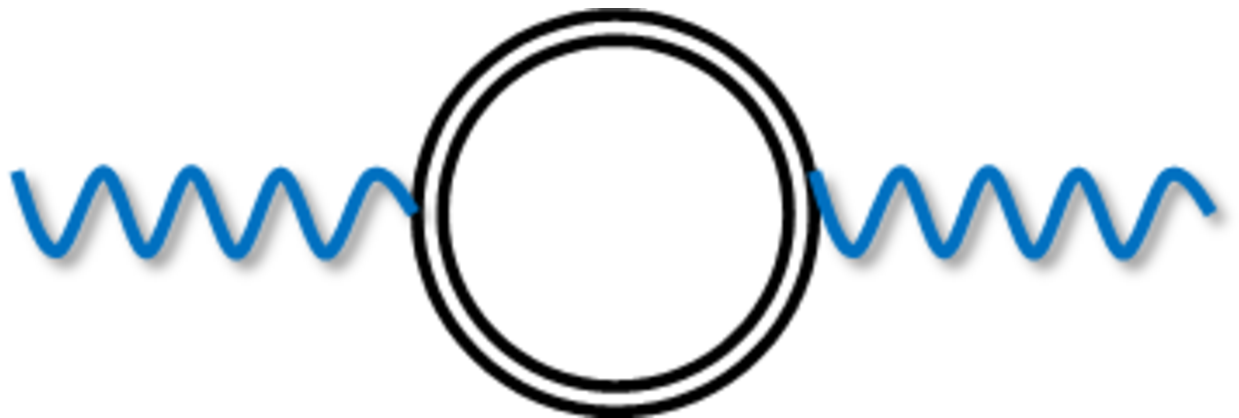}
	\end{center}
\caption{One-loop diagram of the vacuum polarization tensor in a strong magnetic field.}
\label{fig:photon_vp}
\end{minipage}
\end{figure}

We consider a system of massive charged fermions interacting with photons. 
Thus we start with the standard spinor QED Lagrangian:
\begin{eqnarray}
\Lag = \bar \psi \left( i \slashed D - m \right) \psi - \frac{1}{4} F^{\mu\nu} F_{\mu\nu}
\label{eq:Lqed}
\ ,
\end{eqnarray}
where we adopt a convention for the covariant derivative, 
$D^\mu = \partial ^\mu + ie A^\mu(x)$ \cite{PS}. 
Here, ``$e$" and ``$m$" representatively denote charge and mass of a fermion 
($e$ is negative for an electron). 
Since the gauge field $A^\mu$ contains both the external field 
$A_{\rm cl}^\mu$ and dynamical (i.e., propagating) degrees of 
freedom $a^\mu$: $A^\mu=A^\mu_{\rm cl} + a^\mu$, 
the fermion kinetic term in the Lagrangian (\ref{eq:Lqed}) determines 
the coupling of the fermion to the external field. 
Whereas the fundamental QED Lagrangian (\ref{eq:Lqed}) 
describes only the linear interaction at the classical level, 
some of higher-order quantum effects become important 
when they are enhanced due to the strong external field, and give rise to 
 nonlinear interaction among photons. 
For example, when the external field is a strong magnetic field, 
insertion of an external field line gives a factor of ${\cal O}(|eB|/m^2)$ 
and $n$ external field lines, ${\cal O}((|eB|/m^2)^n)$, thereby one has 
to sum up all the diagrams when $B$ is larger than the critical field 
$B\simge B_c\equiv m^2/|e|$ to obtain a ``dressed" fermion propagator. 
This effect is shown in Fig.~\ref{fig:DressedProp}. 
Since this dressed fermion propagator includes all-order contributions 
with respect 
to $eB$, any process involving the dressed fermions becomes nonlinear 
with respect to the external field $B$, and nonperturbative. 
Similarly, one can define the critical electric 
field $E_c\equiv m^2/|e|$ which also indicates breakdown of the ordinary
perturbative expansion. However, electric fields beyond $E_c$ induce 
instability of the vacuum \cite{Dun,Sch}, called the Schwinger mechanism, and will be 
screened by creation of fermion-antifermion pairs. 
On the other hand, the critical magnetic field $B_c$ just indicates 
onset of strong nonlinear effects, and it makes sense to treat magnetic 
fields stronger than the critical one. Thus, as far as one considers 
static magnetic fields, one can discuss very strong nonlinear regime
in QED. This is called the ``nonlinear QED" regime.

If there are several species of fermions $\psi^{(i)}$ with 
different masses $m_{(i)}$ and charges $e_{(i)}$, one can define 
critical magnetic fields $B_c^{(i)}=m_{(i)}^2/|e_{(i)}|$ corresponding 
to each fermion: $B_c^{(1)}<B_c^{(2)}<\cdots$. 
Since magnetic fields can become stronger than the 
(minimum) critical field (unlike the critical electric field beyond which
the Schwinger mechanism occurs), one has to include all the relevant 
fermion degrees of freedom when one treats very strong magnetic field;
typically if $B_c^{(n)}<B<B_c^{(n+1)}$, then at least we need to 
include fermions up to $\psi^{(n)}$.  
In the present paper, however, we provide formula only with 
the lightest fermion. Contributions of heavier fermions 
are easily obtained by appropriate replacements of the electric charge 
and the mass by those of the heavier fermions. Note also that 
contributions from different fermions are all additive in the vacuum 
polarization tensor.

A fundamental key ingredient is the fermion propagator in an external field
$G(x,y|A_{\rm cl}) \equiv \langle 0|{\rm} T \psi(x) \bar \psi(y)|0\rangle$
which includes all the interactions with the external field.  
Recall that the propagator is a Green's function of the Dirac operator. 
If one includes only the external field into the Dirac operator, 
the propagator satisfies  (in the momentum space), 
\begin{eqnarray}
\left( \slashed p - e \slashed A _{\rm cl} - m \right) G(p|A_{\rm cl}) = i
\label{eq:Green}
\, .
\end{eqnarray}
Therefore, one finds 
\begin{eqnarray}
G(p|A_{\rm cl}) &=& \frac{i}{\slashed p - e \slashed A_{\rm cl} - m} 
\label{eq:Gp2}
\\
&=& \frac{i}{\slashed p -m} \sum_{n=0}^\infty 
\left[ \ 
(-ie \slashed A_{\rm cl} ) \frac{i}{\slashed p -m}
\  \right] ^n
\, .
\end{eqnarray}
The last expression allows us to diagrammatically 
depict the nonlinear interaction with the external field as shown in 
Fig.~\ref{fig:DressedProp}.

When one considers the propagation of a photon in a strong background field 
$A^\mu_{\rm cl}$, one uses the fermion propagator (\ref{eq:Gp2}). For 
example, a one-loop contribution of a photon self-energy is computed 
as in Fig.~\ref{fig:photon_vp}. Thus, modification of the fermion propagator 
due to the external field causes a significant influence on the photon 
propagation. 
Representing the photon self-energy, or equivalently the vacuum polarization 
tensor,  in a constant external field as 
$\Pi_{\rm ex}^{\mu\nu}(x-y)$, one finds that 
the kinetic term of the propagating mode $a^\mu$ acquires a quantum effect as, 
$\Lag_A^{\rm loop} = \frac{1}{2} \integ  dy \ a_\mu(x) \ \Pi_{\rm ex}^{\mu\nu} (x-y) \ a_\nu(y)$. 
Then, the Euler-Lagrange equation for $a^\mu$ leads to a modified Maxwell 
equation, 
$\left[ \ q^2 \eta^{\mu\nu} - q^\mu q^\nu - \Pi_{\rm ex}^{\mu\nu} (q) \ \right] a_\nu(q)  =  0$. 
Although the vacuum polarization in the ordinary vacuum does not modify photon propagation owing to the gauge and Lorentz symmetries, 
we show that quantum excitations in an externally applied electromagnetic field behave like electron-hole excitations in dielectric substances \cite{Weisskopf}. 
Indeed, the vacuum polarization tensor corresponds to a response function of 
the Dirac sea to an electromagnetic field induced by an incident photon. 


To compute the vacuum polarization tensor in the external field, 
we use the ``proper-time method" which was developed by 
J.~Schwinger~\cite{Sch}. One can equivalently rewrite 
the dressed propagator (\ref{eq:Gp2}) in a different way as
\begin{eqnarray}
G(p|A_{\rm cl}) =  i\left( \slashed p - e \slashed A_{\rm cl} + m \right) \times \frac{1}{i}
\int_0^\infty \!\! d\hat\tau \ 
{\rm e}^{ i \hat\tau \left\{ (\slashed p -e \slashed A _{\rm cl})^2 - (m^2-i\varepsilon ) \right\} } 
\, ,
\label{eq:GDp}
\end{eqnarray}
where the integral with respect to the ``proper time" $\hat\tau$ is 
convergent  
owing to a prescription by $-i \varepsilon$. Note that $\hat\tau$ 
has dimension of inverse mass squared. Infinite sum with respect to the 
external field is now encoded into the exponential factor in the integrand. 
As summarized in Appendix \ref{sec:PTM}, one can explicitly compute 
$G(p|A_{\rm cl})$ when the external field is constant.

By using the dressed propagator shown in Eq. (\ref{eq:electron}), 
we can now calculate the 1-loop vacuum polarization tensor in the external 
magnetic field (see Fig.~\ref{fig:photon_vp}):
\begin{eqnarray}
i\ \Pi_{\rm ex}^{\mu \nu}(q) = 
(-ie)^2 (-1) \integ \frac{d^4p}{(2\pi)^4} 
{\rm Tr} \Big[ \ \gamma^\mu G(p|A_{\rm cl}) \gamma^\nu 
G(p+q|A_{\rm cl}) \ \Big] 
\label{eq:prop}
\, ,
\end{eqnarray}
where an overall minus sign arises from the fermion loop. 
This is a gauge invariant quantity and should be independent 
of the gauge we adopt.\footnote{In the actual calculation, however, we adopt 
specific gauges. For  computation of the fermion propagator 
$G(p|A_{\rm cl})$, we worked in the Fock-Schwinger gauge for the background 
field $A_{\rm cl}$ (see Appendix \ref{sec:PTM}). As we will see below, 
we then compute the polarization tensor in the covariant gauge which 
fixes the residual gauge symmetry of the dynamical gauge field $a^\mu$.} 
Note that momentum integration gives rise to an ultraviolet divergence, 
because the fermion propagator behaves as $G(p|A_{\rm cl}) \sim p^{-1}$ for 
large $p$.
While we find a quadratic superficial degree of divergence from 
a na\"{i}ve power counting, 
it actually diverges logarithmically owing to a specific tensor structure 
demanded by the Ward identity, $q_{\mu} \Pi_{\rm ex}^{\mu \nu}(q) = 0$. 
We will explicitly show a gauge invariant form 
of the vacuum polarization tensor in the following, 
and comment on renormalization in the end of this section. 

It has been known 
that the vacuum polarization tensor is endowed with 
a gauge-invariant tensor structure in terms of three transverse-projection 
operators satisfying $q_\mu P_i^{\mu\nu}=0$. 
Their explicit forms are given by 
\begin{eqnarray}
&&
\Pi_{\rm ex}^{\mu\nu} (q) = - \Big( \chi_0  P_0^{\mu\nu} + \chi_1 P_1^{\mu\nu} + \chi_2  P_2^{\mu\nu} \Big)\, ,
\label{eq:Pex}
\\
&&
P^{\mu\nu}_0 = q^2 \eta^{\mu\nu}  - q^\mu q^\nu 
\ , \ \ 
P^{\mu\nu}_1 = q^2_\parallel  \eta^{\mu\nu}_\parallel  - q^\mu_\parallel  q^\nu_ \parallel 
\ , \ \ 
P^{\mu\nu}_2 = q^2_\perp \eta^{\mu\nu}_\perp - q^\mu_\perp q^\nu_\perp
\label{eq:Pmn}
\, ,
\end{eqnarray}
where we suppressed arguments of Lorentz-scalar functions 
$\chi_i \ (i=0,1,2)$, of which explicit expressions are immediately 
shown below. We now apply the external magnetic field along the 
third axis of spatial coordinates in the negative direction
so that $eB^3>0$ for an electron ($B^3$ is the third component of the 
magnetic field vector $B^i$). Since the presence of the magnetic field 
specifies a preferred direction, one distinguishes 
longitudinal and transverse directions with respect to the magnetic field.
The metric tensor $\eta^{\mu\nu} = {\rm diag} (1,-1,-1,-1)$ is decomposed into 
longitudinal and transverse subspaces 
$\eta_\parallel^{\mu\nu} = {\rm diag} (1,0,0,-1)$ and 
$\eta_\perp^{\mu\nu} = {\rm diag} (0,-1,-1,0)$, respectively. 
Similarly, longitudinal and transverse momenta are defined as 
$q_\parallel^{\mu} = (q^0, 0, 0, q^3)$ and 
$q_\perp^{\mu} = (0, q^1, q^2, 0)$, respectively.

In the proper-time method (see Appendix \ref{sec:PTM} for more details), 
the polarization tensor (\ref{eq:prop}) has 
integrals with respect to the momentum $p$, and two proper times 
which we denote $\hat\tau_1$ and $\hat\tau_2$ corresponding to each propagator (see Eq.~(\ref{eq:GDp})). 
The integration over $p$ is just a Gaussian integral, 
and can be straightforwardly 
performed. 
Then, we rewrite the remaining integrals by using two dimensionless variables
$\tau = eB ( \hat\tau_1 + \hat\tau_2 )/2 $ and 
$\beta = eB ( \hat\tau_1 - \hat\tau_2 )/\tau$. 
Notice that the scalar functions $\chi_i$ $(i=0,1,2)$  are 
dimensionless, and thus we further introduce three dimensionless variables, 
$\Br = {B}/{B_c}$, $r_\parallel^2 =  \frac{ q_\parallel^2 }{ 4m^2 }$ and 
$r_\perp^2 = \frac{ q_\perp^2 }{ 4m^2 } = - \frac{ | \bq_\perp | ^2 }{ 4m^2 }$
so that the scalar functions $\chi_i$ are expressed as 
\begin{eqnarray}
\chi_i (r_\parallel^2, r_\perp^2; \Br) 
&=& 
\frac{\alpha}{4\pi} \int_{-1}^1 \!\!\! d\beta  \int_0^\infty \!\!\!\! d\tau \ 
\frac{ \Gamma_i (\tau, \beta) }{ \sin \tau } \ 
{\rm e}^{-i u \cos (\beta \tau) } \ 
{\rm e}^{ i \eta \cot \tau } \ 
{\rm e}^{-i\phi_\parallel \tau} \, ,
\label{eq:vp0} 
\end{eqnarray}
where we have introduced two shorthand notations: 
$\eta \equiv - 2 { r_\perp^2 }/{ \Br } $ and $u \equiv { \eta }/{\sin\tau} $,
and lastly $\phi_\parallel$ and $\Gamma_i$ are known functions given by 
the following \cite{Adl1, Tsa, TE, MS76, Urr78} (see also 
Refs.~\cite{DR,DG} for details):
\begin{eqnarray}
&&\phi_\parallel (r_\parallel^2, \Br) 
=  \frac{1}{\Br}\left\{ \ 1 - (1-\beta^2) \ r_\parallel^2 \ \right\}\, , 
\label{eq:phi0} 
\end{eqnarray}
and
\begin{eqnarray}
&&\Gamma_0  (\tau, \beta)
= \cos( \beta \tau) - \beta \sin( \beta \tau) \cot \tau \, ,
\nonumber
\\
&&\Gamma_1  (\tau, \beta) 
= (1-\beta^2) \cos \tau   - \Gamma_0 (\tau, \beta)
\label{eq:Gam0}
\, ,\\
&&\Gamma_2  (\tau, \beta)
= 2 \frac{\cos( \beta \tau ) - \cos \tau}{\sin^2 \tau} - \Gamma_0 (\tau, \beta)
\nonumber
\ \ .
\end{eqnarray}
In Eq.~(\ref{eq:vp0}), the coupling constant in the 
overall factor $\alpha=e^2/4\pi$ comes from propagating photons 
attached to the fermion one-loop, while the others are from 
higher-order effects associated with the external magnetic field 
as is evident from the observation that 
they always appear with the magnetic field as $eB$. 
As will be shown in Sec.~\ref{sec:SeriesExpansion}, 
the scalar function $\chi_1$ is a real-valued function when $\rp \leq 1$, 
reflecting convergence of the double integral in Eq.~(\ref{eq:vp0}), 
and the same holds for the others, $\chi_0$ and $\chi_2$, when $\rp \leq (1+\sqrt{1+2\Br})^2/4$. 
Whereas reliable numerical computation has been performed 
in this kinematical region \cite{KY}, analytic calculation is necessary 
for understanding behavior of the polarization tensor in the 
whole kinematical region. 
In Sec.~\ref{sec:SeriesExpansion}, we will explicitly perform the 
remaining two integrals
to obtain the analytic expression of the polarization tensor.

We emphasize here that we have neither specified any dispersion relation 
for the external photon momentum, nor even assumed that the photon is 
on-shell. We will discuss later that the photon dispersion will be 
determined as a result of interaction with the external magnetic fields. 
In general, there will be a deviation from the massless-type dispersion 
relation indicating that $\rp+\rt \neq 0$ and $\omega\neq |\bq |$. 
Accordingly, the longitudinal and transverse momenta appearing in 
Eqs.~(\ref{eq:vp0}) -- (\ref{eq:Gam0}) should be treated independently 
at this moment.

Recall that only $\chi_0$ survives in the ordinary vacuum 
and is divergent. Divergence of $\chi_0$ is also seen in the presence 
of the magnetic field (while the others are finite). 
This can be explicitly verified as follows. First of all, let us see the 
lower limit of $\tau$ integration in Eq.~(\ref{eq:vp0}). 
If one takes the limit $\tau \rightarrow 0$ in Eq.~(\ref{eq:Gam0}), 
one finds two of them vanish as $\Gamma_1, \Gamma_2 \to 0$ and thus
$\chi_1$ and $\chi_2$ are finite, while $\chi_0$ is logarithmically 
divergent because $\Gamma_0 \to 1-\beta^2$ as $\tau\to 0$. 
In other words, subtraction of $\Gamma_0(\tau,\beta)$ 
in the definition of $\Gamma_1$ and $\Gamma_2$ ensures the finiteness 
of $\chi_1$ and $\chi_2$. Next, one can easily identify 
the origin of the logarithmic divergence by  referring 
the diagrams in Figs.~\ref{fig:DressedProp} and \ref{fig:photon_vp}.
If one resolves the dressed fermion propagator (a double line) in 
Fig.~\ref{fig:photon_vp} into the sum of external magnetic field insertions 
as shown in Fig.~\ref{fig:DressedProp}, 
the first diagram is given by the same vacuum polarization tensor 
as in the ordinary vacuum without any external field attached. 
This diagram has a logarithmic divergence as conventionally known. 
We also notice that this is the only divergent diagram in the series 
because any additional external leg reduces the degree of divergence. 
Therefore, the vacuum polarization tensor (\ref{eq:vp0}) contains the 
same divergence as in the ordinary vacuum, without any additional one. 
As shown in Appendix~\ref{subsec:renorm} in detail, 
we can remove this logarithmic divergence by adopting 
an on-shell (with zero transverse momentum) renormalization 
condition\footnote{
Note that the divergence has not been regularized in 
Eq.~(\ref{eq:vp0}) -- (\ref{eq:Gam0}). 
We will introduce appropriate regularizations depending on analytic 
and numerical calculations, which will be shown in 
Appendix~\ref{subsec:renorm}. } 
such that a finite vacuum polarization tensor 
$\Pexr^{\mu\nu} (q_\parallel, q_\perp ; \Br)$ vanishes at the 
simultaneous limits of on-shell and vanishing field, $\rp,\rt=0$ 
and $\Br = 0$. 
On the basis of this prescription, we define a finite vacuum polarization 
tensor, 
\begin{eqnarray}
\Pexr ^{\mu\nu}  (q_\parallel, q_\perp ; \Br) \equiv 
\Pi_{\rm ex} ^{\mu\nu} (q_\parallel, q_\perp ; \Br) - \Pi_0 ^{\mu\nu} (0)\, ,
\end{eqnarray}
where the vacuum polarization tensor in the ordinary vacuum 
$\Pi_0^{\mu\nu} (q^2)$ is obtained in the vanishing field limit as 
$\lim_{\Br \rightarrow 0} \Pi_{\rm ex} ^{\mu\nu} (q_\parallel, q_\perp ; \Br) 
\rightarrow \Pi_0^{\mu\nu} (q)$. 
We will more specifically discuss these things related to renormalization 
in Appendix \ref{subsec:renorm}.


\section{Optical property of the vacuum with magnetic fields} 
\label{sec:optics}

Before we go into the detailed calculation of 
$\chi_i$, let us discuss observable effects of nonzero $\chi_i$, which 
does not require explicit expressions of $\chi_i$. In particular, we 
introduce the notions of dielectric constants and refractive indices that 
are frequently used in context of optics of dielectric substances
and semiconductors.

Propagating photons in the ordinary vacuum have two transverse 
oscillating modes with the dispersion relation of the massless type, 
$\omega^2 = \left| \bq \right|^2$, which does not vary even if 
quantum corrections are included, as long as the gauge and Lorentz 
symmetries are preserved.
The presence of an externally applied magnetic field however 
breaks the Lorentz symmetry, and the dispersion relation of a photon 
is subject to modification which leads to two intriguing phenomena: 
two propagating (physical) modes can have refractive indices different from 
unity, and a photon is able to decay into a fermion-antifermion pair if 
its energy is large enough.
In this section, we show kinematical aspects of these effects 
on the basis of the Maxwell equation modified by a nonlinear interaction 
with the external magnetic field.

In the presence of an external magnetic field, the Maxwell equation 
for the dynamical photon field $a_\mu$ is modified as 
\begin{equation}
\left[ \ q^2 \eta^{\mu\nu} - \left( 1-\frac{1}{\xi_g} \right) q^\mu q^\nu 
- \Pi_{\rm ex}^{\mu\nu}(q_\parallel, q_\perp ; \Br) \ \right] a_\nu(q)  =  0\, , 
\end{equation}
where $\xi_g$ is a parameter in the gauge fixing term, 
$\Lag_{\rm GF} = - \frac{1}{2 \xi_g} (\partial^\mu a_\mu)^2$. 
Substituting the vacuum polarization tensor (\ref{eq:Pex}),  
one finds 
\begin{eqnarray}
\left[ \  (1+\chi_0) P_0^{\ \mu\nu} + \chi_1 P_1^{\ \mu\nu} + \chi_2 P_2^{\ \mu\nu} + \frac{1}{\xi_g}  q^\mu q^\nu  \ \right] a_\nu(q) = 0
\, .
\label{eq:Maxwell_2}
\end{eqnarray}
Below we explain in a qualitative way how the vacuum birefringence 
and on-shell photon decay follow from this modified Maxwell equation. 
To this end, explicit expressions of the 
scalar functions $\chi_i$ are not important.

In order to identify two physical modes in the modified Maxwell 
equation (\ref{eq:Maxwell_2}), we need to `diagonalize' the matrix 
equation by selecting appropriate vectors associated with the direction 
of the magnetic field. First of all, it should be noticed that one can 
define three independent vectors that are orthogonal to the momentum
of the propagating photon $q^\mu$. 
In addition to the invariant tensors $\eta^{\mu\nu}$ and 
$\epsilon^{\mu\nu\rho\sigma}$, one can utilize the field-strength tensor 
$F^{\mu\nu}$ and the photon momentum $q^\mu$ to construct independent vectors. 
Namely, one finds that 
the following four vectors $v_{(\lambda)}^\mu$ $( \lambda = 0,\cdots,3)$ are orthogonal 
among each other \cite{Sha}:
\begin{eqnarray}
&&
v_{(0)}^\mu = q^2 F^{\mu\nu} F_{\nu\rho} q^\rho - 
( q^\nu F_{\nu\rho} F^{\rho \sigma} q^\sigma) q^\mu\, ,
\nonumber
\\
&&
v_{(1)}^\mu = F^{\mu\nu} q_\nu 
\ , \ \ 
v_{(2)}^\mu = \tilde F^{\mu\nu} q_\nu 
\ , \ \ 
v_{(3)}^\mu = q^\mu  \, ,
\label{eq:vec1}
\end{eqnarray}
where the dual field-strength tensor is defined as 
$\tilde F^{\mu\nu} 
= \frac{1}{4} \epsilon^{\mu\nu\rho\sigma} F_{\rho\sigma}$ 
with the completely antisymmetric tensor, $\epsilon^{0123} = 1$. 
Indeed, one can show the orthogonal relations \footnote{
In general, when electric ($\bE$) and magnetic ($\bB$) fields 
are present, one finds $v_{(\lambda)}^\mu \ v^{}_{(\sigma)\mu}=0$ except for 
$ v_{(1)}^\mu \ v^{}_{(2)\mu}= - q^2 (\bE \cdot \bB)$.
Therefore, the orthogonality among all the vectors holds 
when $\bE\cdot \bB=0$ and thus $\bE =0$ (our case).}: 
$v_{(\lambda)}^\mu \ v^{}_{ (\sigma) \mu} = 0$ for any 
$\lambda$ and $\sigma(\neq \lambda)$.

Assuming that there is only a magnetic field and it 
is oriented to the third direction of the spatial coordinates, the set 
of vectors (\ref{eq:vec1}) simplifies to 
\begin{eqnarray}
&&
v_{(0)}^\mu = B^2 \ (q_\perp^2 q_\parallel^\mu - q_\parallel^2 q_\perp^\mu)\, ,
\nonumber
\\
&&
v_{(1)}^\mu = B \ \tilde q_\perp^\mu
\ , \ \ 
v_{(2)}^\mu = B \ \tilde q_\parallel^\mu 
\ , \ \ 
v_{(3)}^\mu = q^\mu 
\, ,
\label{eq:vec2}
\end{eqnarray}
where $B$ is a magnitude of the external magnetic field. For convenience,
we have introduced two momentum vectors 
$\tilde q_\parallel^\mu = (q^3, 0, 0, q^0)$ and $\tilde q_\perp^\mu = (0, q^2, -q^1,0)$. 
They are orthogonal to all the previously introduced momenta, 
$q^\mu$, $q_\parallel^\mu$ and $q_\perp^\mu$. 
Therefore, the vectors $\tilde q_\parallel^\mu$ and $\tilde q_\perp^\mu$ 
span a space complementary to the space spanned by 
$q_\parallel^\mu$ and $q_\perp^\mu$.

Notice that the projection operators in the modified Maxwell equation 
(\ref{eq:Maxwell_2}) are orthogonal to the momentum vectors, 
$q^\mu$, $q_\parallel^\mu$ and $q_\perp^\mu$, namely, 
$q_\mu P_i^{\ \mu\nu}=q_{\parallel\mu} P_i^{\ \mu\nu}
=q_{\perp\mu} P_i^{\ \mu\nu}=0$. This fact suggests that one can 
represent the projection operators in a different way by using the 
vectors $\tilde q_\parallel^\mu$ and $\tilde q_\perp^\mu$. 
Indeed, two of the projection operators are 
expressed as $P_1^{\ \mu\nu} = - \tilde q_\parallel^\mu \tilde q_\parallel^\nu $ and $ P_2^{\ \mu\nu} = \tilde q_\perp^\mu \tilde q_\perp^\nu $. 
Therefore, one finally finds that 
the projection operators and thus the modified Maxwell equation
are `diagonalized' as follows:
\begin{eqnarray}
&&P_0^{\ \mu\nu} = q^2 \ ( \pi_{(0)}^\mu \pi_{(0)}^\nu + \pi_{(1)}^\mu \pi_{(1)}^\nu + \pi_{(2)}^\mu \pi_{(2)}^\nu  )
\ , \ \ 
P_1^{\ \mu\nu} = q_\parallel^2 \   \pi_{(2)}^\mu \pi_{(2)}^\nu  
\ , \ \ 
P_2^{\ \mu\nu} = q_\perp^2 \   \pi_{(1)}^\mu \pi_{(1)}^\nu  
\, ,
\label{eq:proj}\\
&&M^{\mu\nu}(q)a_\nu 
= 
\sum_{\lambda=0}^{3}{\cal M}_{(\lambda)}(q) 
\pi^{\mu}_{(\lambda)}\pi^\nu_{(\lambda)} \, a_\nu
=0\, ,
\end{eqnarray} 
where we have rescaled $v_{(\lambda)}^\mu$ so that they have unit 
norms as\footnote{Our definition of $\pi^\mu_{(\lambda )}$ contains 
somewhat confusing notation wherein the subscripts of the last two 
projection operators in Eq.~(\ref{eq:proj}) 
oppositely correspond to those of the vectors on the right-hand sides. 
However, we keep this notation because the subscripts originally 
defined in Eq.~(\ref{eq:vec1}) give natural ordering (i.e., 
$P_1$ is given by $\pi_{(1)}$, etc) in case of 
electric fields.} 
\begin{equation}
\pi_{(\lambda)}^\mu \equiv 
\frac{v_{(\lambda)}^\mu}{\sqrt{v_{(\lambda)}^\nu 
v^{}_{(\lambda)\nu}}}\qquad 
(\lambda = 0,1,2,3)\, .\label{pol_vectors}
\end{equation}
and the coefficients ${\cal M}_{(\lambda)}$ are expressed in 
terms of $\chi_i$. 
One can easily find the eigen-modes of the modified Maxwell equation 
by expanding the dynamical photon field in terms of 
$\pi_{(\lambda)}^\mu$ and assuming a plane wave solution: 
\begin{eqnarray}
a^\mu(q) = N \sum_{\lambda=0}^3 \pi_{(\lambda)}^\mu 
\ {\rm e}^{-i(\omega t - \bq \cdot \bx) }
\ \ ,
\label{eq:Aprop}
\end{eqnarray}
where $q^\mu = (\omega, \bq)$ and 
$N$ is a dimensionful normalization constant. Notice again that we have 
not specified the dispersion relation which 
provides a relation between $\omega$ and $\bq$, 
and this is what we suppose to obtain 
by solving the modified Maxwell equation.
Indeed, one finds that linear independence of 
$\pi_{(\lambda)}^\mu$ leads to a set of eigenvalue-equations,
\footnote{We note that the same results can be obtained from 
the pole position of a photon propagator 
in the magnetic field (see Appendix \ref{sec:Pprop}).}  
\begin{eqnarray}
\left\{
\begin{array}{l}
\{ \ ( 1+\chi_0 ) \ q^2  \ \} \ \pi_{(0)}^\mu = 0\, ,
\\
\{ \ ( 1+\chi_0 ) \  q^2 + \chi_2  \ q_\perp^2  \ \} \ \pi_{(1)}^\mu = 0\, ,
\\
\{ \ ( 1+\chi_0 ) \  q^2 + \chi_1  \ q_\parallel^2  \ \} \ \pi_{(2)}^\mu = 0\, ,
\\
\xi_g^{-1} \ q^2 \ \pi_{(3)}^\mu = 0
\ \ ,
\end{array}
\right.
\label{eq:eigen}
\end{eqnarray}
which immediately yields the dispersion relation of a photon: 
\begin{equation}
\epsilon  = \frac{ | \bq |^2 }{  \omega^2}\, . 
\label{eq:eps_def}
\end{equation} 
Here we have defined a dielectric constant $\epsilon$ 
which persists unity in the ordinary vacuum.

It is easily seen that the first and fourth equations lead to  
dispersion relations of the massless type, $\omega^2=|\bq|^2$, 
with the speed of light in the ordinary vacuum ($\epsilon = 1$),
as long as neither $(1+\chi_0) = 0$ nor $\xi_g^{-1}=0$. 
However, as explained in Appendix \ref{sec:EM},  
these two modes turn out 
to be unphysical (see Eqs.~(\ref{unphysical0}) and (\ref{unphysical3})) 
and thus we do not discuss them below.

In contrast, the second and third equations yield 
nontrivial dispersion relations of physical modes. 
Without loss of generality, 
we suppose that a photon is propagating in a plane ($y=0$) 
spanned by the first and third directions of spatial coordinates, 
and that the external magnetic field is oriented to the (negative) third 
direction. 
The momentum vector $q^\mu$ is now represented as 
$q^\mu = ( \omega, |\bq| \sin (\pi -\theta), 0, |\bq| \cos (\pi -\theta)  )$, 
where $\theta$ denotes an angle between 
the direction of the external magnetic field and the momentum of 
a propagating photon. Then, the second and third equations yield 
the vacuum dielectric constants $\epsilon_\perp$ and $\epsilon_\parallel$, 
respectively: 
\begin{eqnarray}
&&\epsilon_\perp = \frac{1+\chi_0}{1+\chi_0+\chi_2 \sin^2 \theta }  \qquad 
{\rm for} \quad  \pi_{(1)}^\mu\, , 
\label{eq:eps1bperp}
\\
&&\epsilon_\parallel = \frac{1+\chi_0+\chi_1}{1+\chi_0+\chi_1 \cos^2 \theta } 
\qquad  {\rm for} \quad  \pi_{(2)}^\mu
\, .
\label{eq:eps1bparallel}
\end{eqnarray}
The same expressions were obtained in Ref.~\cite{MS76} 
in the radiation gauge, while we have obtained them in the covariant gauge. 
The agreement of two calculations in two different gauges indicates  
gauge invariance of the dielectric constant, as expected. 
Whereas we have not specified the explicit form of $\chi_i$,
two constants $\epsilon_\perp$ and $\epsilon_\parallel$ are in general 
not equal to unity and different from each other.\footnote{
In the ordinary vacuum, only $\chi_0$ is nonzero. 
Even with nonzero $\chi_0$, the above 
expression reduces to $\epsilon_\perp=\epsilon_\parallel=1$ 
when $\chi_1 = \chi_2 = 0$, as expected. 
}
Namely, we have found that the externally applied magnetic field gives rise 
to two distinct dielectric constants depending on the polarization directions. 
This immediately implies that we can define two distinct refractive indices. 
Therefore, we call this phenomenon {\it vacuum birefringence} after a similar 
phenomenon in dielectric substances. 
Notice also that, due to the violation of the Lorentz symmetry by an 
external magnetic field, 
the dielectric constants explicitly depend on the zenith angle with respect to the magnetic field direction. 
Nevertheless, the system maintains a boost invariance 
in the direction of the constant external magnetic field, 
and thus photon propagations in the directions at $\theta =0, \pi$ are 
special. Substituting these angles into Eqs.~(\ref{eq:eps1bperp}) and 
(\ref{eq:eps1bparallel}), 
we find that both of the dielectric constants become unity 
$\epsilon_\perp(\theta=0,\pi)=\epsilon_\parallel(\theta=0,\pi) =1$, 
as a consequence of the boost invariance.

Let us further assume that the scalar functions $\chi_i$ have 
imaginary parts. Then, the dielectric constants also have imaginary 
parts: $\epsilon = \epsilon_{\rm real} + i \, \epsilon_{\rm imag}$. 
We will see later in Sec. \ref{sec:SeriesExpansion} that imaginary parts 
indeed appear under some kinematical conditions. Since the 
dielectric constants and 
refractive indices are related with each other via  $n^2 = \epsilon$,
we can similarly define real and imaginary parts of the refractive 
indices:\footnote{Since there are two polarization modes, one can define two 
different refractive indices $n_\parallel$ and $n_\perp$, both of which 
can be complex quantities.}
\begin{equation}
n = n_{\rm real} + i \, n_{\rm imag}
\ \ ,
\end{equation}
where one can easily find 
($| \epsilon |=\sqrt{ \epsilon_{\rm real} ^2 + \epsilon_{\rm imag} ^2 }$)
\begin{eqnarray}
\label{eq:ref_real}
&& n_{\rm real} \ = 
\ \frac{1}{\sqrt{2}} \sqrt{  | \epsilon | + \epsilon_{\rm real}  }\, ,  
\\
&& n_{\rm imag} \ = 
\ \frac{1}{\sqrt{2}} \sqrt{  | \epsilon | - \epsilon_{\rm real}  }  
\, .
\label{eq:ref_imag}
\end{eqnarray}
While a purely real dielectric constant yields a real refractive 
index\footnote{
Inserting Eqs.~(\ref{eq:eps1bperp}) and (\ref{eq:eps1bparallel}) into this expression 
and then expanding them with respect to $\chi_0 \sim 0$, $\chi_1 \ll 1 $, 
and $\chi_2 \ll 1$, we obtain approximate forms shown in the literature, 
$n_{\perp} = 1 - (\chi_2/2)  \sin^2 \theta$ and 
$n_{\parallel} = 1 + (\chi_1/2) \sin^2 \theta$. 
However, one should bear in mind that 
$\chi_0$, $\chi_1$ and $\chi_2$ can be even divergent and complex 
depending on kinematical variables, 
as we will see in Sec.~\ref{sec:SeriesExpansion}. 
} 
$n_{\rm real}=\sqrt{\epsilon_{\rm real}}$,
a purely imaginary dielectric constant leads to an equal magnitude of 
real and imaginary parts, $n_{\rm real} = n_{\rm imag} = 
 \sqrt{ | \epsilon | /2} $ with $|\epsilon |=|\epsilon_{\rm imag}|$. 
In Appendix \ref{sec:EM}, we have explained that real and 
imaginary parts of the two refractive indices $n_\parallel$ and $n_\perp$ 
in general provide distinct phase velocities and extinction coefficients of 
physical modes $\lambda=1,2$ in Eq. (\ref{eq:Aprop}), respectively.

It should be noticed and taken care of carefully that 
the right-hand sides of Eqs.~(\ref{eq:eps1bperp}) and (\ref{eq:eps1bparallel}) 
depend on the dielectric constants $\eperp$ and $\epara$, respectively, 
through the photon momenta $\rp$ and $\rt$. 
According to the definition of the dielectric constant (\ref{eq:eps_def}), 
those momenta are written as $\rp = \ome^2 (1- \epsilon \, \cos^2 \theta)$ 
and $\rt = - \epsilon \, \ome^2 \sin^2 \theta$, 
where we introduced a normalized photon energy $\ome^2 = \omega^2/(4m^2)$. 
Thus, if one wants to obtain the dielectric constant $\epsilon$ as a function of the photon energy $\ome$, 
Eqs.~(\ref{eq:eps1bperp}) and (\ref{eq:eps1bparallel}) have to be solved with respect to $\epsilon$. 
Physically, this structure reflects effects of a back-reaction appearing as 
screening of an incident photon field by an induced vacuum polarization, 
which should be taken into account self-consistently when the magnitude of the polarization becomes large. 
As mentioned above, the dielectric constants are in general complex, 
and thus damping of the incident photon field, 
due to the decay into a fermion-antifermion pair, should also be
treated self-consistently. 
Therefore, we need to simultaneously solve two sets of coupled equations 
(obtained from Eqs.~(\ref{eq:eps1bperp}) and (\ref{eq:eps1bparallel})) for 
the real and imaginary parts of $\epsilon_\perp$ and $\epsilon_\parallel$
in a self-consistent way. There are indeed kinematical regions
where these procedures play an important role to correctly obtain the 
dielectric constant. We will explicitly demonstrate it in the next 
paper~\cite{HI2}. 


\section{Analytic evaluation of the vacuum polarization tensor} 
\label{sec:SeriesExpansion}


In the previous sections, we have discussed in a formal way 
that a nonzero vacuum polarization tensor in a magnetic field 
necessarily gives rise to the vacuum birefringence. In order to know 
how large the effect is and how it depends on kinematical conditions, 
we have to explicitly evaluate the scalar functions $\chi_i \ (i=0,1,2)$ 
of which formal expressions were already given in 
Eqs.~(\ref{eq:vp0}) -- (\ref{eq:Gam0}). 
Notice that this representation contains somewhat complicated integration 
with respect to $\beta$ and $\tau$, corresponding to the difference and 
the average of two proper-time variables. In fact, this complexity 
has prevented previous studies from complete analytical understanding 
of the vacuum birefringence \cite{Adl1,TE,MS76}, 
and even from performing numerical computation 
except in a limited kinematical region \cite{KY}. 
We will, in this section, perform the 
integration {\it analytically} by rewriting the integrand into a double 
infinite series of familiar functions, and show that the infinite
series indeed has a physical meaning. 
We will see that the scalar functions $\chi_i$ and thus the 
refractive indices are sensitively affected by microscopic 
structure of the fermion spectrum in a magnetic field, i.e., the 
Landau levels formed by the fermions. Therefore, 
as in optics for materials, 
macroscopic electromagnetism of the vacuum should be elucidated by the 
microscopic dynamics of the Dirac sea which is excited by an incident 
photon field.

\subsection{Computing the scalar functions $\chi_i$}

\label{subsec:exp}

Before we present the procedures how to analytically perform the integrals, 
let us briefly point out technical difficulties in the 
integral representation, Eqs.~(\ref{eq:vp0}) -- (\ref{eq:Gam0}). 
Note that each integrand in $\chi_i$ contains an exponential function 
whose argument is given by trigonometric functions. This factor causes 
strong oscillation of the integrand composed of arbitrarily higher harmonics, 
because $k$-th power of the trigonometric function that appears 
in a Taylor expansion of the exponential factor is composed of 
higher harmonics up to $k$ times the fundamental 
frequency. 
For example, the factor ${\rm e}^{i\eta \cot \tau}$ contains
a term proportional to $\cos^3 \tau$ at $k=3$ in the Taylor expansion, 
which is composed of up to the 3rd higher harmonics as 
$\cos^3 \tau = \{ 3 \cos \tau + \cos (3 \tau) \} /4$.  
In general, this strong oscillation even cannot be periodic 
due to mixing of two fundamental 
periods, $2\pi$ and $2\pi / \beta$. 
These complicacies, directly or indirectly, cause 
difficulties in analytic and numerical evaluations of the integral. 
In fact, as we will see later, results of the integration have singular behaviors 
at some kinematical conditions, which may invalidate numerical
approaches.

The first thing to overcome this situation 
is to rewrite the integrand so that the exponential function 
does not contain trigonometric functions of the 
fundamental 
period, $2\pi/\beta$ (Step I). 
We will still have an exponential function with a trigonometric 
function in the argument which is however composed of a unique 
fundamental 
period, $2\pi$. 
We can rewrite it by a series expansion so that the exponential 
function does not contain any trigonometric function (Step II). 
The resultant integrand will allow us to easily perform analytic 
integration (Step III).
Below, we explain these procedures step by step.

\subsubsection{Step I: Partial wave decomposition}

Consider the functions in the integrand in Eq.~(\ref{eq:vp0}) 
that depend on trigonometric functions with a fundamental period $2\pi/\beta$. Namely, 
${\rm e}^{-iu\cos (\beta\tau)}$, 
$\cos (\beta\tau){\rm e}^{ -iu\cos (\beta\tau)}$, and 
$\cos (\beta\tau){\rm e}^{ -iu\cos (\beta\tau)}$. We can equivalently rewrite 
these functions by using the partial wave decomposition \cite{BK}.
By adopting the formulas given in Appendix \ref{sec:pwd}, 
we obtain the following\footnote{
We just use the partial wave decomposition as a mathematical tool. 
Later, the index $n$ is interpreted as a physical quantity, 
but not an angular momentum.}:
\begin{eqnarray}
\chi_i 
= 
\frac{\alpha}{4\pi} \int_{-1}^1 \!\!\! d\beta \int_0^\infty \!\!\!\! d\tau \ 
\sum_{n=0}^\infty \ ( 2 - \delta_{n0} )
\frac{ \gamma_i^{(n)} (\tau, \beta) }{ \sin \tau } \ 
{\rm e}^{ i\eta \cot \tau } \, 
{\rm e}^{-i( \phi_\parallel -n\beta )\tau} \, ,
\label{eq:vp1} 
\end{eqnarray}
with $\gamma_i^{(n)}(\tau,\beta)$ defined by
\begin{eqnarray}
&&\gamma_0^{(n )} (\tau, \beta)
\equiv  
\frac{1}{2} \left\{ \,  I_{n+1}(-iu) + I_{n-1}(-iu) \, \right\} - n\beta \ \eta^{-1} I_n(-iu) \cos\tau\, , \nonumber
\\
&&\gamma_1^{(n )} (\tau, \beta)
\equiv  (1-\beta^2)  I_n(-iu) \cos\tau    - \gamma_0^{(n )} (\tau, \beta)\, , 
\label{eq:Gam1}
\\
&&\gamma_2^{(n )} (\tau, \beta)
\equiv 
\sin^{-2} \! \tau
\left\{ \, I_{n+1}(-iu) + I_{n-1}(-iu) - 2  I_n(-iu) \cos\tau \, \right\}  
- \gamma_0^{(n)} (\tau, \beta)
\, ,\nonumber
\end{eqnarray}
where  $I_n(-iu)$ is the modified Bessel function of the first kind. 
Except for the last trivial exponential factor 
${\rm e}^{-i( \phi_\parallel -n\beta )\tau}$ in Eq.~(\ref{eq:vp1}), the 
integrand now oscillates with the fundamental period $2\pi$.
Note that $I_n(-iu)$ also oscillates with the fundamental period $2\pi$ because 
its argument is proportional to cosecant as, $-iu = -i \eta \csc \tau $.



\subsubsection{Step II: Infinite series expansion}

We are able to further simplify the integrand by using the following 
technique. We first introduce a variable $z$ defined by 
\begin{eqnarray}
z \equiv \exp ( -2i \tau )\, .\nonumber
\end{eqnarray}
Then,
a product of the exponentiated cotangent and the modified Bessel function
in Eqs.~(\ref{eq:vp1}) and (\ref{eq:Gam1}) are rewritten as 
$
{\rm e}^{ i \eta\cot\tau }  \ I_n(-iu) 
=  {\rm e}^{-\eta}  {\rm e}^{ - z\frac{2\eta}{1-z} } \ 
I_n  ( \frac{ 2 ( \eta^2 z)^{\frac{1}{2}} }{1-z} ) 
$. 
We use the following formula which is deduced from the 
relation between confluent geometrical functions (see Sec.~10.12 in Ref.~\cite{Erd}):
\begin{eqnarray}
\exp \left( -z \frac{ x+y }{ 1-z } \right) I_n \left( \frac{2 (xyz)^{\frac{1}{2}} }{1-z} \right)
= (1-z) \, (xyz)^{\frac{n}{2}} \sum_{\ell=0}^{\infty} \frac{ \ell! }{\Gamma( \ell + n + 1 )} 
\ L_\ell^n(x) L_\ell^n(y) \ z^\ell
\, ,
\label{eq:IL}
\end{eqnarray}
where $L^n_\ell(x)$ and $\Gamma(x)$ are 
the associated Laguerre polynomial and the Gamma function, respectively. 
Notice that this relation disentangles 
the function with a nontrivial $z$-dependence (left-hand side) into a sum of 
polynomials of $z$ (right-hand side), which is quite
useful for the present purpose. Namely, the resulting 
integrand can be now written only in terms of simple 
exponentials of $\tau$. 
Taking parameters in the above formula as $x = y = \eta$ and 
using $ (1-z)^{-2} = \sum_{i=0}^\infty \sum_{j=0}^\infty z^{i+j} = \sum_{k=0}^\infty (k+1) \ z^k$, 
we find that the double integrals are simplified to either of
the following representations:
\begin{eqnarray}
F_{\ell}^n (r_\parallel^2, \Br) 
&\equiv & 
\frac{i}{\Br}
\int_{-1}^1 \!\! d\beta \int _0^\infty \!\! d\tau
\ {\rm e}^{-i  \left( \phi_\parallel + 2\ell -n\beta + n \right) \tau  }\, ,
\label{eq:F0}
\\
G_{\ell}^n (r_\parallel^2, \Br) 
&\equiv& 
\frac{i}{\Br}
\int_{-1}^1 d\beta \int _0^\infty d\tau
\ \beta \, 
{\rm e}^{-i  \left( \phi_\parallel + 2\ell -n\beta + n \right) \tau  }\, ,
\label{eq:G0}
\\
H_{\ell}^n (r_\parallel^2, \Br) 
&\equiv& 
\frac{i}{\Br}
\int_{-1}^1 d\beta \int _0^\infty d\tau 
\ \beta^2 \, 
{\rm e}^{-i  \left( \phi_\parallel + 2\ell -n\beta + n \right) \tau  }
\label{eq:H0}
\, .
\end{eqnarray}
Apart from these integrals, we have nontrivial coefficients which 
contain the associated Laguerre polynomials as we will 
explicitly show in the final expression below. 
Here we just emphasize that, thanks to the formula (\ref{eq:IL}),
the longitudinal ($r_\parallel^2$) dependence of 
$\chi_i(r_\parallel^2, r_\perp^2)$ appears only in the functions 
$F,G$ and $H$, while the transverse ($r_\perp^2$) 
dependence are factorized and appears in the other parts.


\subsubsection{Step III: Performing double integrals}

The integrals in Eqs.~(\ref{eq:F0}) -- (\ref{eq:H0}) are simple enough 
and we can carry out integration analytically. 
However, it should be noticed that these integrals contain singular behaviors, 
which appear when the exponential factor becomes unity due to a vanishing argument. 
Depending on the photon momentum $\rp$, 
there arises such a divergent contribution within the integral region, $\beta \in [-1,1]$. 
A simple way to obtain the integrals with respect to any $\rp$ is
 provided by the technique of analytic continuation. 
First, assuming that the rescaled photon momentum $\rp$ satisfies 
$\phi_\parallel(r_\parallel^2, \Br) + \left( 2\ell -n\beta + n \right) > 0 $, 
we perform convergent integration by rotating the contour in the complex $\tau$-plane downward. 
Then, analytic continuation allows us to inspect the analytic property 
afterward. 
At this moment, the $\tau$-integral is carried out as follows: 
\begin{eqnarray}
F_{\ell}^n (r_\parallel^2, \Br) 
&=&
\int_{-1}^1  
\frac{ d\beta }{ \ r_\parallel^2 \beta^2 - n \Br \beta + (1-r_\parallel^2) + (2\ell+ n) \Br \ }
\equiv I_{\ell \Delta}^n (r_\parallel^2)
\ \ .
\end{eqnarray}
The integral with respect to $\beta$ simply provides a difference 
$ I_{\ell \Delta}^n (r_\parallel^2) = I_{\ell +}^n(r_\parallel^2) - I_{\ell -}^n (r_\parallel^2)$ 
between the following functions:
\begin{eqnarray}
I_{\ell \pm}^n (r_\parallel^2) \equiv \frac{2}{\sqrt{4ac-b^2}}   \arctan \left(  \frac{b\pm2a}{\sqrt{4ac-b^2}}  \right)
\ \ .
\label{eq:Ipm}
\end{eqnarray}
We have introduced parameters $a=r_\parallel^2$, $b= - n \Br $ and 
$c = ( 1 - r_\parallel^2 ) + ( 2\ell +  n ) \Br  $, 
and the integral with respect to $\beta$ has been carried out in a regime, 
$4ac-b^2 > 0$. Note that the condition for $\rp$ assumed above involves this regime. 
Similarly, the remaining other two integrals are performed as
\footnote{
Although the expressions of $G_{\ell}^n (r_\parallel^2, \Br) $ 
and $H_{\ell}^n (r_\parallel^2, \Br) $ apparently look singular 
at $\rp = 0$, one can check that both of them are regular there. 
} 
\begin{eqnarray}
G_{\ell}^n (r_\parallel^2, \Br) 
&=&
\frac{1}{ 2 r_\parallel^2 }
\left[ \ \Xi_\ell^n (\Br) + n \Br \ 
I_{\ell \Delta}^n (r_\parallel^2) \ \right]\, ,
\\
H_{\ell}^n (r_\parallel^2, \Br) &=&
\frac{1}{r_\parallel^2}
\left[ \ 
2 + \frac{n\Br}{2 r_\parallel^2} \ \Xi_\ell^n (\Br) 
+ \frac{1}{ 4r_\parallel^2 } \left\{ \ ( b^2 -4 ac ) +  ( n\Br)^2  \ \right\}
I_{\ell \Delta}^n (r_\parallel^2) 
\ \right]
\, ,
\end{eqnarray}
where we have introduced a real constant $\Xi$ (i.e., independent of 
kinematical variables): 
$$
\Xi_\ell^n (\Br)  \equiv 
\ln \left| \frac{ 1 + 2 \ell \Br }{ 1 + 2 ( \ell + n ) \Br } \right| 
= \ln \left| \frac{ m^2 + 2 \ell eB }{ m^2 + 2 ( \ell + n ) eB } \right|\, .
$$
We have performed the integrals under the condition 
$4ac-b^2 > 0$. However, the obtained results can be analytically continued 
to the regime $4ac-b^2 < 0$, where the function 
$I_{\ell \Delta}^n (r_\parallel^2)$ will become complex as we 
will see later.

\subsubsection{Final results}

Now that we have performed all the integrals, we find the analytic expression
of $\chi_i$ $(i=0,1,2)$. In order to present the results in a compact form, 
let us define the coefficients $C_\ell^n (\eta)$ expressed by the associated 
Laguerre polynomials\footnote{
We employ a convention for the associated Laguerre polynomial 
defined by series: 
$L_\ell^n(\eta) = \frac{{\rm e}^\eta \eta^{-n}}{\ell!} \frac{d^\ell \ }{d\eta^\ell} \left( {\rm e}^{-\eta} \eta^{\ell+n} \right) 
= \displaystyle \sum_{r=0}^{\ell} \  {}_{\ell+n} \mathrm{C} {}_{\ell-r} \  \frac{(-\eta)^r}{r!}  
\, .$}:
\begin{eqnarray}
C_\ell^n (\eta) \equiv 
{\rm e}^{ - \eta }  
\frac{\ell !}{ ( \ell + n ) ! } \, \eta^n  \big[ L_\ell^{n} (\eta) \big]^2
\, . \label{eq:C}
\end{eqnarray}
Then, the coefficients $\chi_i$ of the vacuum polarization tensor 
(\ref{eq:Pex}) are finally represented as the infinite sum of known 
functions as 
\begin{eqnarray}
\chi_i 
& = &
\frac{ \alpha \Br }{ 4\pi } \, 
\sum_{n=0}^{\infty} \ (2-\delta_{n0} ) 
\left[ 
\ 
\sum_{\ell=0}^{\infty} \Omega_{\ell \, i}^{n (0)}
+
\sum_{\ell=1}^{\infty} \Omega_{\ell \, i}^{n (1)}
+
\sum_{\ell=2}^{\infty} \Omega_{\ell \, i}^{n (2)}
\ 
\right]\, ,
\label{eq:vp2} 
\end{eqnarray}
where we have introduced three functions 
$\Omega_{\ell \, i}^{n (0)}$, $\Omega_{\ell \, i}^{n (1)}$, and
$\Omega_{\ell \, i}^{n (2)}$, as we define below.
 
For $i=0$, they are defined as 
\begin{eqnarray}
&&
\Omega_{\ell \, 0}^{n (0)} =   
(1-\delta_{n0}) \ C_\ell^{n-1} (\eta)  \  F_{\ell}^{n} (\xi, \Br)
- n \eta^{-1} \ C_\ell^n  (\eta) \  G_{\ell}^{n}  (\xi, \Br)\, ,
\nonumber
\\
&&\Omega_{\ell \, 0}^{n (1)}
=
(1+\delta_{n0}) \ C_{\ell-1}^{n+1} (\eta)  \  F_{\ell}^{n} (\xi, \Br)
- \ 
n \eta^{-1} \ C_{\ell-1}^n (\eta) \  G_{\ell}^{n}  (\xi, \Br)\, ,
\nonumber
\\
&&\Omega_{\ell \, 0}^{n (2)} =0\, .
\end{eqnarray}
Note that the functions $F$ and $G$ (and also $H$) 
depend on $\rp$ only through the combination, $ \xi \equiv 2\rp/\Br$. 
Thus, we have written them as $F(\xi,\Br)$ and $G(\xi,\Br)$. 
Similarly, for $i=1$, we have 
\begin{eqnarray}
&&\Omega_{\ell \, 1}^{n (0)}
 = 
C_\ell^n (\eta)  \left\{  
F_\ell^n (\xi,\Br) - H_\ell^n (\xi,\Br)
 \right\}
- \Omega_{\ell \, 0}^{n (0)}\, ,\nonumber \\
&&\Omega_{\ell \, 1}^{n (1)} 
= 
C_{\ell-1}^n (\eta)  \left\{  
F_\ell^n (\xi,\Br) - H_\ell^n (\xi,\Br)
 \right\}
- \Omega_{\ell \, 0}^{n (1)}\, ,\nonumber \\
&&\Omega_{\ell \, 1}^{n (2)} = 0\, ,
\end{eqnarray}
and finally for $i=2$,
\begin{eqnarray}
&&\Omega_{\ell \, 2}^{n (0)} = - \Omega_{\ell \, 0}^{n (0)}\, ,\nonumber\\
&&\Omega_{\ell \, 2}^{n (1)} =  \ D_{\ell } ^{n (1)}  (\eta) \  
F_{\ell}^{n}  (\xi, \Br) - \Omega_{\ell \, 0}^{n (1)}\, ,\nonumber\\
&&\Omega_{\ell \, 2}^{n (2)} =  \ D_{\ell } ^{n (2)}  (\eta) \  
F_{\ell}^{n} (\xi, \Br)\, .
\end{eqnarray}
In the last expression, we introduced coefficient functions $D^{(1)}$ and 
$D^{(2)}$ defined by 
\begin{eqnarray}
&&D_{\ell } ^{n (1)}  (\eta) = -8 
\displaystyle
\sum_{\lambda=0}^{\ell-1} \ \left( \ell - \lambda \right)  
\left\{(1-\delta_{n0}) \ C_\lambda^{n-1} (\eta) - C_\lambda^n (\eta) 
\right\}\, ,
\label{eq:D1}
\\
&&D_{\ell } ^{n (2)}  (\eta) =-8 
\displaystyle
\sum_{\lambda=0}^{\ell-2} \  \left( \ell -\lambda - 1 \right)  
\left\{  (1+\delta_{n0}) \ C_\lambda^{n+1}  (\eta)  - C_\lambda^n  (\eta) 
 \right\}\, .
 \label{eq:D2}
\end{eqnarray}
These are the final analytic results represented as an infinite 
series of known functions. For fixed $n$ and $\ell$, each term in the 
expansion is finite (except for some special kinematical points as we will
discuss soon). 
The ultraviolet divergence of the function $\chi_0$ appears only 
after one takes the infinite summation (see Appendix~\ref{subsec:renorm}).

Figure~\ref{fig:C} shows $\ell$-dependence 
of the coefficient $\C(\eta)$ in Eq.~(\ref{eq:C}) 
at fixed $\eta$ and $n$. 
The associated Laguerre polynomial provides an oscillation, 
and, in case of finite $n$, a ratio of the factorials strongly 
suppresses $\C(\eta)$ for small $\ell$. 
In Appendix~\ref{sec:formulae}, we discuss the 
behavior of $\C(\eta)$ in the limit of 
large $\ell \gg 1$. 
By using the limiting form (\ref{eq:Clim}) of $\C(\eta)$ at 
large values of $\ell$ with fixed $\eta$ and $n$, 
one can put the upper bound for $\C(\eta)$: 
\begin{eqnarray}
0
\leq
\C(\eta)
\leq 
\frac{1}{\pi \sqrt{ \eta\ell \  } }\, {\rm e}^{ -\frac{n+1}{2\ell} } 
\ \ \ ( \ell \gg 1)
\label{eq:Cabs}
\, ,
\end{eqnarray}
Black dashed 
line shows the upper bound of $\C(\eta)$ for 
$n = 0$, and the other two cases are also bounded at large $\ell$ outside 
the plot range. 
While the upper bound in Eq.~(\ref{eq:Cabs}) indicates that 
$\C(\eta)$ for large $\ell$ converges to zero by an inverse-square-root, 
this suppression is not sufficiently strong to guarantee a convergence of the infinite sum (\ref{eq:vp2}).  
Indeed, we find a logarithmic divergence of the function $\chi_0$ by carrying out the infinite sum. 
However, the logarithmic divergence has been foreseen {\it a priori} to appear 
in terms of the ultraviolet divergence mentioned below Eq.~(\ref{eq:Gam0}), 
which should be removed by an appropriate prescription (see 
Appendix~\ref{subsec:renorm}).

\begin{figure}[t]
  \begin{center}
   \includegraphics[width=0.75\hsize]{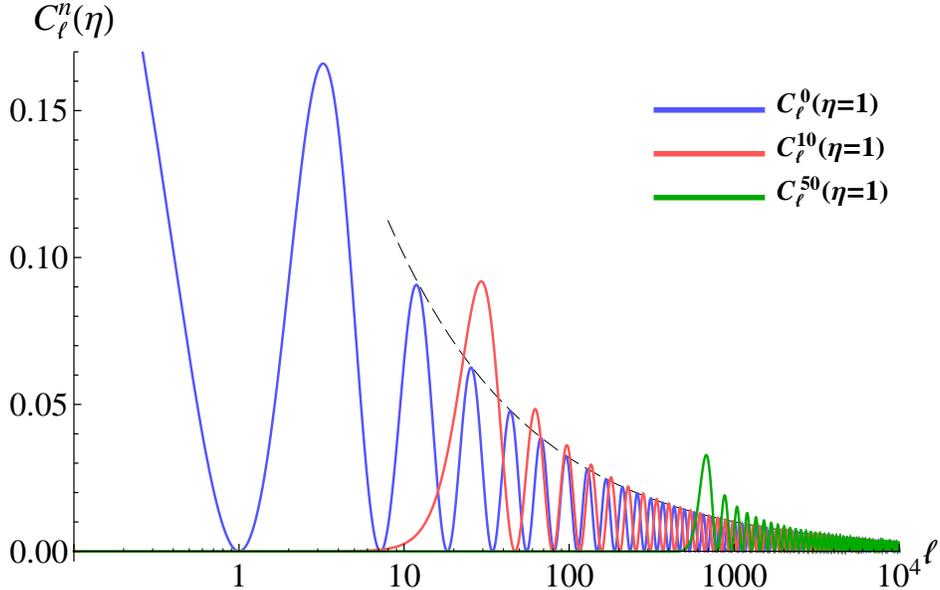}
  \end{center}
\caption{
Coefficient $C_\ell^n(\eta)$ at fixed $n$ and $\eta$: 
$C_\ell^n(\eta)$ at discrete $\ell$ takes a value 
on curves plotted against continuous $\ell$. 
A dashed line shows the upper bound for the coefficient 
at $n=0$ that is valid at large $\ell \gg 1$. 
Owing to Eq.~(\ref{eq:Clim}), other two curves are also bounded at large 
$\ell$ outside the plot range. 
Note that the coefficient has a finite value at $\ell=0$ given by 
$C_0^n (\eta) = {\rm e}^{-\eta} \eta^n / n ! $. 
}
\label{fig:C}
\end{figure}

\subsection{Physical meaning of the results: Landau levels}

\label{sec:kinematics}

The use of infinite series expansion was motivated rather by 
technical reasons for making the double integrals simpler. 
However, as we will see below, the 
indices $\ell$ and $n$ in the expansions have well-defined 
physical meaning: they are related to the Landau levels of 
the fermion-antifermion pair which appears in the one-loop diagram 
shown in Fig.~\ref{fig:photon_vp}.

First of all, it should be noticed that 
an important kinematical property of the scalar function $\chi_i$, 
with respect to $\rp$, 
is essentially determined by the function $I^n_{\ell \Delta}(\rp)$ 
which is the only one possible source of an imaginary part. 
Therefore, we investigate the structure of $I^n_{\ell \Delta}(\rp)$ 
in order to understand kinematics in the presence of an external 
magnetic field, and to identify the physical meaning of the indices 
$\ell$ and $n$.

A kinematical property of $I^n_{\ell \Delta}(\rp)$ is specified 
by a discriminant appearing in Eq~(\ref{eq:Ipm}): 
\begin{eqnarray}
{\mathcal D}&\equiv & b^2-4ac\nonumber\\
&=& ( - n \Br )^2 - 4\rp \left[(1-\rp)+(2\ell +n)\Br \right]\, .
\label{discriminant}
\end{eqnarray}
Depending on the sign of $\mathcal D$, 
arguments of the arctangent in $I^n_{\ell \Delta}(\rp)$ are either real or pure imaginary. 
Solving an equality, $\mathcal D = 0$, we find 
boundaries of the momentum region specified by a pairwise solution, 
\begin{eqnarray}
\rp =  \frac{1}{4} \left\{ 
\sqrt{ 1 + 2 \ell \Br } \pm \sqrt{ 1 + 2 ( \ell + n ) \Br }  \right\}^2 
\equiv s_\pm^\idx
\ \ .
\label{eq:s_pm}
\end{eqnarray} 
If the photon momentum resides in the regime 
$s_-^\idx < r_\parallel^2 < s_+^\idx $ where ${\cal D}>0$, 
the function $I_{\ell \Delta}^n (r_\parallel^2)$ is obviously 
a real-valued function. On the other hand, 
in the other regimes, $r_\parallel^2 < s_-^\idx $ and 
$s_+^\idx < r_\parallel^2$ where ${\cal D}<0$, 
it possibly becomes a complex function. However, 
we show in Appendix \ref{sec:atan} that 
$I_{\ell \Delta}^n (r_\parallel^2)$ has an imaginary part 
only if the photon momentum resides in the higher regime, 
$s_+ ^\idx< \rp$. 
In other words, it does not have an imaginary part in the 
complementary regime, $\rp < s_+^ \idx $ which includes 
$\rp < s_-^\idx $. 
After careful investigation shown in Appendix \ref{sec:atan}, 
we obtain a piecewise representation as, 
\begin{eqnarray}\hspace*{-5mm}
I_{\ell \Delta}^n (r_\parallel^2) 
= 
\left\{
\begin{array}{l}
\frac{ 1 }{ \sqrt{ (r_\parallel^2 - s_-^\idx )  (r_\parallel^2 - s_+^\idx ) } } 
\cdot \frac12
\ln \left|  \frac{ a-c - \sqrt{b^2-4ac} }{ a-c + \sqrt{b^2-4ac} }  \right|
\ \ \hspace{38mm}
( r_\parallel^2<s_-^\idx )
\\
\frac{ 1 }{ \sqrt{  \left\vert (r_\parallel^2 - s_-^\idx )  (r_\parallel^2 - s_+^\idx ) \right\vert  } } 
\left[  
\arctan \left(  \frac{b+2a}{\sqrt{4ac-b^2}}  \right)
 -
\arctan \left(  \frac{b-2a}{\sqrt{4ac-b^2}}  \right)
 \right] \quad
( s_- ^\idx < r_\parallel^2 < s_+ ^\idx )
\\
\frac{ 1 }{ \sqrt{ (r_\parallel^2 - s_-^\idx )  (r_\parallel^2 - s_+^\idx )  } } \cdot \frac12 
\left[ \ 
\ln \left|  \frac{ a-c - \sqrt{b^2-4ac} }{ a-c + \sqrt{b^2-4ac} }  \right|
+  2 \pi i \ 
\right]
\ \ \hspace{20mm}
( s_+^\idx < r_\parallel^2)\, .
\end{array}
\right.
\label{eq:I}
\end{eqnarray}

An imaginary part in $I_{\ell \Delta}^n (r_\parallel^2) $ results in 
the imaginary part of the scalar functions $\chi_i$, and of 
the vacuum polarization tensor $\Pi_{\rm ex}^{\mu\nu}(q) $. 
Consequently, the dielectric constants and thus refractive indices 
have imaginary parts as alluded 
in Eqs.~(\ref{eq:eps1bperp}) -- (\ref{eq:ref_imag}). 
They represent damping of photon propagation in the presence of 
an external magnetic field. As foreseen owing to the optical theorem, 
this damping is caused by decay into a fermion-antifermion pair. 
Thus, we interpret $\rp=s_+^\idx$ as the threshold momentum of photon decay, 
and notice that there are infinite number of threshold momenta 
as the indices $\ell$ and $n$ run from zero to infinity. 

In terms of dimensionful quantities, the threshold condition $\rp=s_+^\idx$ 
for the photon decay is expressed as 
\begin{eqnarray}
q_\parallel^2 = \left\{ \sqrt{  m^2 + 2 \ell eB \ } + \sqrt{  m^2 + 2 ( \ell + n ) eB \ }  \right\}^2
\, .
\label{eq:thrs}
\end{eqnarray}
Recall that the energy of a charged particle in a magnetic field is 
$\varepsilon_n(p_z)=\sqrt{m^2+p_z+2neB}$. 
This immediately implies that 
the right-hand side of Eq.~(\ref{eq:thrs}) exactly agrees with 
the invariant mass of a fermion-antifermion pair carrying quantized 
transverse momenta in the magnetic field and vanishing longitudinal 
momenta along the external field. 
The integers $\ell$ and $\ell+n$ specify the Landau levels. 

Reflecting the boost invariance of the system along the external constant 
magnetic field, the left-hand side of Eq.~(\ref{eq:thrs}) has a boost 
invariant form $q_\parallel^2$. 
It indicates that the decay condition should be invariant 
in a class of Lorentz frames which are connected by the boost along 
the external field, 
because the propagating photon receives an influence from 
the magnetic field of the same configuration. 
Owing to the boost invariance, 
it is possible, without varying the configuration of the external field, 
to take the Lorentz frame 
in which the longitudinal momentum of 
the photon vanishes,\footnote{If one wants to perform an equivalent 
analysis in a Lorentz frame disconnected to this class, 
one has to take into account an electric field orthogonal to 
the magnetic field.}  $q_{(0) z} = 0$. 
In this Lorentz frame, we have $q_\parallel^2 = \omega_{(0)}^2 - q_{(0)  z}^2 
= \omega_{(0)}^2$, and the condition (\ref{eq:thrs}) represents 
nothing but the smallest photon energy to produce a fermion-antifermion 
pair in the Landau levels $\ell$ and $\ell + n$ with vanishing longitudinal 
momenta. 

For each $\ell$ and $n$, the function $I^n_\ell(\rp)$ has a single 
threshold momentum beyond which a photon can decay into a fermion-antifermion 
pair with the Landau levels specified by $\ell$ and $\ell + n$. 
Since the coefficients $\chi_i$ and thus the vacuum polarization tensor are 
given as the infinite sum of $I^n_\ell(\rp)$ over the indices $\ell$ and $n$, 
there are infinite number of threshold momenta in the photon momentum $\rp$. 
Beyond the each threshold momentum, 
the vacuum polarization tensor has an imaginary part, 
indicating a branch cut continuously running on the real axis in the complex 
$q_\parallel^2$-plane. 
The half line of the cut and its starting point at the threshold 
correspond to the continuous and vanishing longitudinal momenta of 
a real pair excitation from the vacuum, respectively.

In order to confirm our interpretation of the Landau levels, let us 
look at the contribution from the lowest Landau levels, $\ell = n =0$. 
Taking $\ell = n =0$ in Eq.~(\ref{eq:vp2}), we find that the coefficients $\chi_i$ 
of the vacuum polarization tensor survive only for $i=1$. Namely, they read 
$ \chi_0 = \chi_2 = 0$, and 
\begin{eqnarray}
\chi_1(\rp, \rt ; \Br) = \frac{\alpha \Br }{4\pi } \ {\rm e}^{-\eta}  
\times \frac{1}{r_\parallel^2} \left\{ I_{0 \Delta}^0 (r_\parallel^2) - 2  
\right\}
\, ,
\label{eq:chi1_LLL}
\end{eqnarray}
where we have used $L_0^0(\eta) = 1$. 
Depending on the values of 
$r_\parallel^2$ (the boundaries are $s_-^{00}=0$ and $s_+^{00}=1$  
(or, $q_\parallel^2 = 4m^2$), see Eq.~(\ref{eq:s_pm})), 
the piecewise expression of $I_{0 \Delta}^0 (r_\parallel^2)$ follows 
from Eq.~(\ref{eq:I}) as 
\begin{eqnarray}
I_{0 \Delta}^0 (r_\parallel^2)
= 
\left\{
\begin{array}{l}
\frac{1}{ \sqrt{ r_\parallel^2(r_\parallel^2-1) } }
\ln \left| 
\frac{ r_\parallel^2 - \sqrt{  r_\parallel^2(r_\parallel^2-1)  } }{ r_\parallel^2 + \sqrt{  r_\parallel^2(r_\parallel^2-1)  }} 
\right|
\ \hspace{24mm} ( r_\parallel^2  < 0  )
\\ 
\frac{2}{\sqrt{ r_\parallel^2( 1 - r_\parallel^2) } }
 \arctan \left\{ \frac{r_\parallel^2}{\sqrt{r_\parallel^2(1-r_\parallel^2)} } \right\}
\ \qquad\ \qquad( 0< r_\parallel^2 < 1 )
\\ 
\frac{1}{ \sqrt{ r_\parallel^2(r_\parallel^2-1) } }
\left[ \ 
\ln \left| 
\frac{ r_\parallel^2  - \sqrt{r_\parallel^2(r_\parallel^2-1) } }{ r_\parallel^2  + \sqrt{r_\parallel^2(r_\parallel^2-1) }} 
\right|
+  \pi i
\ \right]
\ \qquad ( 1 < r_\parallel^2  )
\end{array}
\right.
\label{eq:I_LLL}
\, .
\end{eqnarray}
Notice that this result does not depend on $\Br$, reflecting the fact that 
the threshold condition $q_\parallel^2 = 4m^2$ is independent of $\Br$.

In principle, one could compute the vacuum polarization tensor 
by an alternative method. 
By using fermion propagators which are decomposed into the Landau 
levels \cite{CEO90,GMS96}, the one-loop diagram can be 
calculated order-by-order with respect to the Landau levels. 
The outcome would however look rather complicated 
compared with our calculation in the proper-time method. 
Nevertheless, as far as the lowest Landau level is concerned, 
a simple calculation leads to an analytic expression of 
the vacuum polarization tensor at this level. 
This calculation was recently carried out in Ref.~\cite{Fuk}, 
and the result is exactly the same as above. Therefore, our analytic 
results correctly reproduce the lowest Landau level contribution, suggesting 
the legitimacy of our interpretation of the Landau levels.

\begin{figure}
	\begin{center}\hspace*{-5mm}
		\includegraphics[scale=0.7]{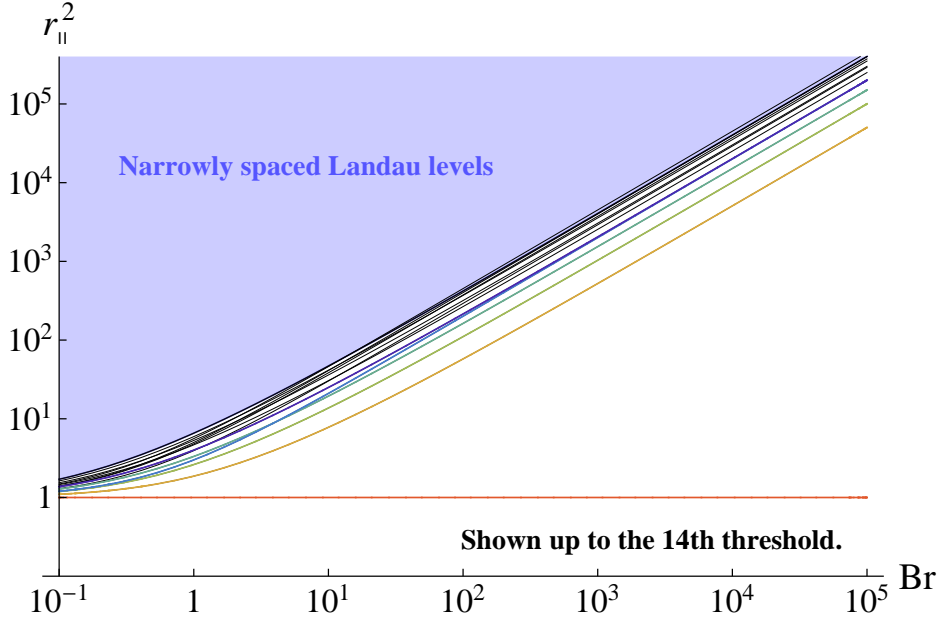}
	\end{center}
\caption{
Threshold structure of photon momentum as a function of $\Br$. Curves are 
positions of thresholds $\rp=s_+^\idx$. We have shown threshold lines from 
the lowest (horizontal line at $\rp=1$) up to the 14th. 
With increasing $\Br$, 
the lowest threshold is isolated from higher levels more distantly, 
and thus the lowest Landau level approximation is justified 
in a wider kinematical region below the second threshold. 
}
\label{fig:Landaulevels}
\end{figure}

Note that the vacuum birefringence is specified by the external 
photon momentum and the magnitude of the external magnetic field. 
Then, 
we notice that the lowest Landau level contribution is not necessarily 
enough for the calculation of 
the dielectric constants and refractive indices of energetic photons. 
Figure~\ref{fig:Landaulevels} shows variation of the threshold structure 
with increasing magnetic field $\Br$. The vertical line corresponds 
to the longitudinal photon momentum squared $\rp=q_\parallel^2/(4m^2)$, 
and curves show locations of 
the thresholds $\rp=s_+^\idx$. 
We have shown only up to the 14th threshold. The high momentum region 
above the 14th threshold line (i.e., the shaded region) is filled 
with narrowly spaced many curves. This figure visualizes that 
there is certainly a region (i.e., low momentum and strong field) 
where the lowest Landau level approximation is appropriate. 
However, if a photon momentum $\rp$ approaches higher levels, 
the levels close to the photon momentum 
would provide dominant contributions and thus 
higher-level contributions have to be incorporated in calculations 
even in strong magnetic field limits. 
With the results of a series representation in Eq. (\ref{eq:vp2}), 
important contributions from individual Landau levels can be 
systematically taken into account in calculating the vacuum polarization 
tensor.

\subsection{Behavior of the coefficients $\chi_i$ at the thresholds}

Let us carefully see the behavior of the function 
$I_{\ell \Delta}^n (r_\parallel^2)$ at the boundaries $\rp=s_\pm ^\idx$,
which should be essentially the same as those of the coefficients $\chi_i$. 
First of all, given the explicit form of $I_{\ell \Delta}^n (r_\parallel^2)$
in Eq.~(\ref{eq:I}), one might expect that singularities  could appear 
at both of the boundaries due to the inverse-square-root. But in fact, 
one finds that $I_{\ell \Delta}^n (r_\parallel^2)$ is smooth at the lower 
boundary $\rp=s_-^\idx$ while it has a discontinuity with divergence at the 
higher boundary $\rp=s_+^\idx$ (see Appendix \ref{sec:limit} for more 
details).

\begin{figure}[t]
	\begin{center}
		\includegraphics[scale=0.68]{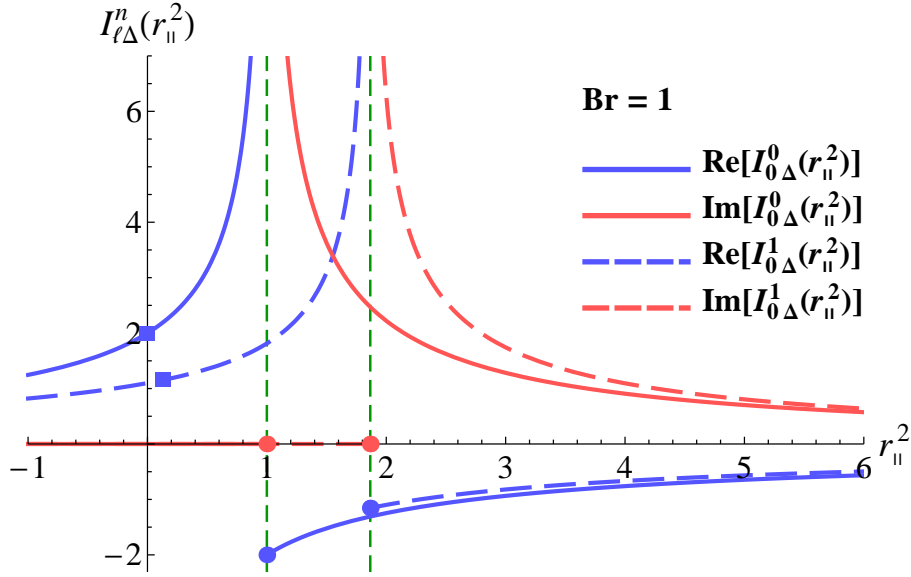}
	\end{center}
\caption{
The function $ I_{\ell \Delta}^n (\rp) $ is plotted against $\rp$ 
for different values of $n$ and $\Br=1$. 
Blue (red) solid and dashed lines are real 
(imaginary) parts of $ I_{\ell \Delta}^n (\rp) $ for $n=0$ and 1, 
respectively. Filled squares (circles) on the lines are lower (higher) 
thresholds $\rp=s_-^{\idx}$ ($\rp=s_+^{\idx}$).}
\label{fig:I}
\end{figure}

As $\rp$ approaches the lower boundary $\rp=s_-^\idx$ from below and above, 
one finds that the values of $I_{\ell \Delta}^n (\rp)$ are finite and coincide 
with each other:
\begin{eqnarray}
  \lim_{\rp \rightarrow s_-^\idx -0} I_{\ell \Delta}^n (\rp) 
= \lim_{\rp \rightarrow s_-^\idx +0} I_{\ell \Delta}^n (\rp) 
= \left. \frac{ 2 }{ c-a } \right|_{\rp = s_-^\idx}
\hspace{-0.5cm}
= 2 \rho (\ell,n; \Br)
\label{eq:limit_r_lower}
\, ,
\end{eqnarray}
where a limiting value is given by 
$\rho (\ell,n; \Br) = \{ ( 1+2\ell\Br ) ( 1+2 (\ell+n) \Br ) \}^{-{1}/{2}}$. 
Note that this is a real number as we commented before. On the other hand,
if $\rp$ goes beyond the higher boundary $\rp=s_+^\idx$ (threshold for the 
photon decay), the function 
 $I_{\ell \Delta}^n (\rp)$ has an imaginary part. Then we find that 
both the real and imaginary parts of $I_{\ell \Delta}^n (\rp)$ show 
singular behavior at the higher boundary $\rp=s_+^\idx$. The real part goes
\begin{eqnarray}
&&\lim_{\rp \rightarrow s_+^\idx -0}  I_{\ell \Delta}^n (\rp) \ \ 
=\lim_{\rp \rightarrow s_+^\idx -0}
\frac{ \lambda(\rp)}{\sqrt{ s_+^\idx - \rp }}
= + \infty\, , \label{eq:limit_r_higher_real-}\\
&& 
\lim_{\rp \rightarrow s_+^\idx +0} \ {\rm Re} \left[ I_{\ell \Delta}^n (\rp) \right]
= \left. \frac{ 2 }{ c-a } \right|_{\rp = s_+^\idx}
\hspace{-0.5cm}
= - 2 \rho (\ell,n; \Br)
\, ,
\label{eq:limit_r_higher_real+}
\end{eqnarray}
where the coefficient $\lambda (\rp) \equiv \pi (\rp -s_-^\idx)^{-{1}/{2}}$ 
takes a finite positive value at the threshold, $\lambda (s_+^\idx) = \pi \rho^{\frac{1}{2}} (\ell,n; \Br) $. 
On the other hand, the imaginary part diverges as $\rp$ approaches 
$\rp = s_+^\idx$ from above, while it is just zero from below: 
\begin{eqnarray}
&&\lim_{\rp \rightarrow s_+^\idx -0}{\rm Im} 
\left[ I_{\ell \Delta}^n (\rp) \right] =0\, ,\\
&&\lim_{\rp \rightarrow s_+^\idx +0}{\rm Im} 
\left[ I_{\ell \Delta}^n (\rp) \right] 
=
\lim_{\rp \rightarrow s_+^\idx +0}\frac{\lambda(\rp)}{\sqrt{\rp - s_+^\idx}} 
= + \infty \, .
\label{eq:limit_i}
\end{eqnarray}
Such singular behavior appears only at the higher boundary $\rp=s_+^\idx$
for each number of $\ell$ and $n$. 
The finite value $\rho (\ell,n; \Br) $ and the coefficient $\lambda(\rp)$,
as well as the position of the singularity, 
depend on the indices $\ell$, $n$, and the magnitude of 
the magnetic field $\Br$, 
but the global structure of the function $ I_{\ell \Delta}^n (\rp) $ is 
common for any $\ell$, $n$ and $\Br$. This is shown in Fig.~\ref{fig:I}.

\begin{figure}[p]
\hspace{\hsize}
\\
	\begin{center}\hspace*{5mm}
		\includegraphics[scale=0.75]{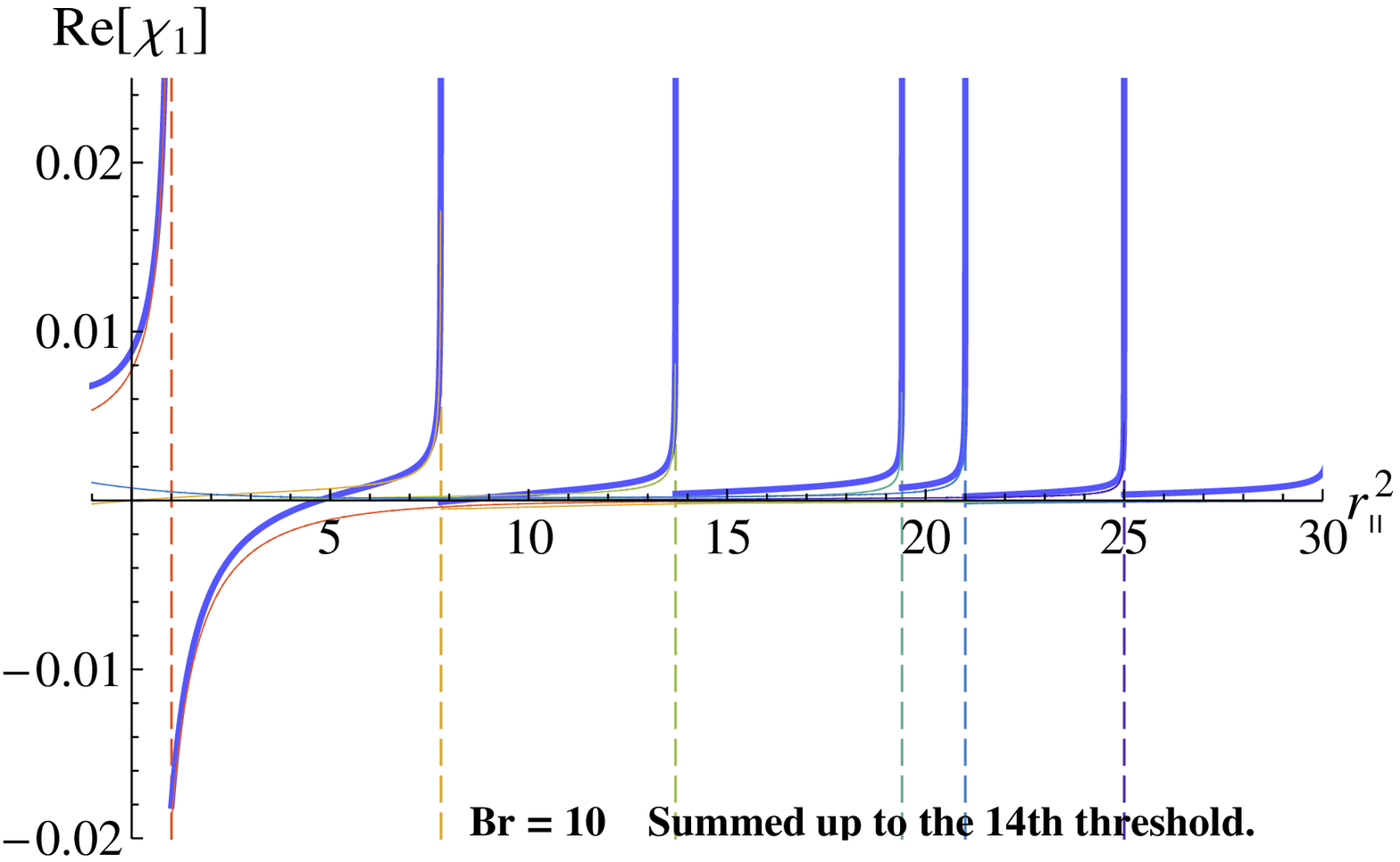}
	\end{center}
\hspace{\hsize}
\\
	\begin{center}\hspace*{-5mm}
		\includegraphics[scale=0.75]{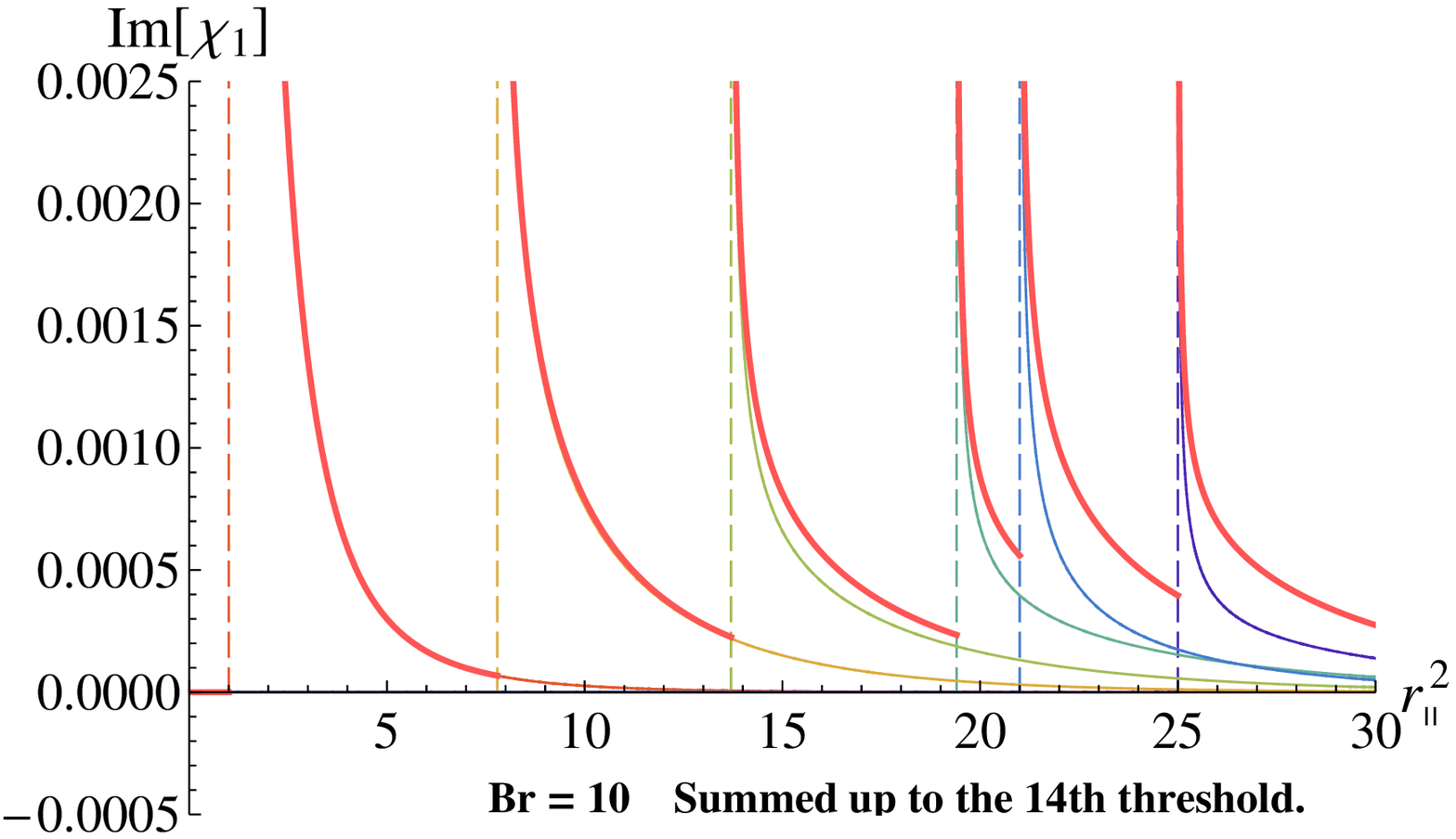}
	\end{center}
\caption{Real and imaginary parts of the coefficient $\chi_1$ as functions of $\rp$ for $\Br=10$. 
Shown are the results obtained with the first 14th terms. 
While thin colored lines indicate individual contributions 
from Landau levels shown with the same colors in Fig.~\ref{fig:Landaulevels}, 
a thick line indicates a resultant quantity after the summation. 
}
\label{fig:chi1}
\end{figure}

Recall that the coefficients $\chi_i$ are functions of 
$I_{\ell \Delta}^n (\rp)$ and thus they will show the same threshold 
behavior. Since $\chi_i$ are obtained after the summation over all
the Landau levels specified by $\ell$ and $n+\ell$, 
and each $I_{\ell \Delta}^n (\rp)$ has a 
singular point for each $n$ and $\ell$, $\chi_i$ have divergences 
at infinitely many 
thresholds. 
In Fig.~\ref{fig:chi1}, we show the real and imaginary parts of 
$\chi_1$ as functions of $\rp$ for $\Br=10$. 
Among infinitely many terms, we have 
summed from the first term up to the 14th term. 
Spikes correspond to the thresholds for different values of $\ell$ and $n$.
For example, the first and second thresholds appear at 
$\rp=s_+^{00}=1$ and $\rp=s_+^{01}=(1+\sqrt{1+2\Br})^2/4\sim 8$ for $\Br=10$. 
It is important to notice that these divergences are 
harmless in the dielectric constants $\epsilon_{\perp},\ 
\epsilon_{\parallel}$ and the refractive indices $n_{\perp},\ n_{\parallel}$. 
This is easily understood from the explicit representation of the dielectric
constants,  Eqs.~(\ref{eq:eps1bperp}) and (\ref{eq:eps1bparallel}). 
First, the function  $I_{\ell \Delta}^n (r_\parallel^2)$  enters all the coefficients $\chi_i\, (i=0,1,2)$, and thus three of the coefficients have 
divergences of the same order at exactly the same momenta. Second, 
these coefficients appear
both in the denominators and numerators of Eqs.~(\ref{eq:eps1bperp}) and 
(\ref{eq:eps1bparallel}). Therefore, the singularities of 
$I_{\ell \Delta}^n (r_\parallel^2)$ bring the same divergences in the 
denominators and numerators, and are 
cancelled to give finite values 
of the dielectric constants and the refractive indices.

We also notice that magnitudes of the dielectric constants, 
in the vicinity of a threshold specified by $\ell$ and $n$, 
are governed by the prefactors of the terms proportional 
to $I_{\ell \Delta}^n (r_\parallel^2)$ in $\chi_i$, 
because they dominate all the other finite terms 
including the terms proportional to 
$I_{\ell^\prime \Delta}^{n^\prime} (r_\parallel^2)$ with 
indices $\ell'$, $n'$ different from $\ell$, $n$.

\section{Summary and prospects}


In this paper, we have presented analytic representation of the 
vacuum polarization tensor in a strong magnetic field, which is 
an indispensable quantity for the description of photon's birefringence. 
This was made possible by the use of infinite series expansion 
applied to the exponentiated trigonometric functions, 
whose integer indices have been later physically interpreted 
as the Landau levels of a fermion-antifermion pair. 
The infinite sum corresponds to the contributions from all the Landau levels 
to the vacuum polarization tensor. 
We have carefully investigated 
the singularity structure of the vacuum polarization tensor at 
an infinite number of the thresholds beyond which 
a photon decays into a fermion-antifermion pair. 

Having obtained the analytic expressions of the 
vacuum polarization tensor, we can proceed to computation of 
the dielectric constants $\epsilon_{\perp,\parallel}$ and refractive indices 
$n_{\perp,\parallel}$ by using Eqs.~(\ref{eq:eps1bperp}) and (\ref{eq:eps1bparallel}). 
However, this does not go straightforwardly 
as we have already emphasized in the last paragraph of Sec.~\ref{sec:optics}. 
We need to solve Eqs.~(\ref{eq:eps1bperp}) and (\ref{eq:eps1bparallel}) 
self-consistently with respect to the dielectric constants. 
In Ref.~\cite{KY}, such self-consistent solutions were numerically obtained 
when the photon momentum is smaller than 
the lowest threshold $\rp<1$. 
The result shows that one of the refractive indices starts to 
deviate from unity as the magnetic field strength increases 
beyond the critical one $\Br>1$, 
and approaches a limiting value, depending on propagation angle, 
in the extremely strong field limit $\Br\gg 1$.  
Our calculation of course precisely reproduces the results of Ref.~\cite{KY}, 
and can go beyond it. 
However, we have to be very careful 
when we investigate a region where the deviation of the dielectric constant 
becomes sizeable. 
In the next paper \cite{HI2}, we will inspect 
the self-consistent solutions 
when the photon momentum $\rp$ increases across the threshold.

The general expression obtained in this work beyond the threshold 
has opened new possibilities to study a wide variety 
of phenomena which contain quite different scales. 
Recall 
the threshold structure shown in Fig.~\ref{fig:Landaulevels}, 
and note that the kinematical region discussed in Ref.~\cite{KY} 
covers 
a small region below the first threshold (horizontal line at $\rp=1$). 
However, if one wants to apply to relativistic heavy-ion collision, 
photons generally have large momenta, 
and thus we have to take into account the effects of many thresholds. 
Also, in application to the physics in magnetars, one would be interested in 
the interplay and competition among photon splitting, 
vacuum birefringence and photon decay \cite{Adl70,Bar}. 
To investigate these intriguing phenomena, 
we need to accurately describe the structure of the photon spectrum 
with 
the thresholds taken into account. 
In the next paper \cite{HI2}, we will discuss 
how and to what extent the dielectric constants and the refractive indices are 
influenced by the singular behavior of the vacuum polarization tensor at the threshold. 
We will show that the location of the threshold itself has to be obtained 
in a self-consistent way. 
In particular, we will investigate behavior of refractive indices 
incorporating effects of thresholds, 
which can be studied within the lowest Landau level approximation.


\section*{Acknowledgements}
This work was supported by Korea national research foundation under 
grant number KRF-2011-0030621, and also partially supported by 
``The Center for the Promotion of Integrated Sciences (CPIS)" of Sokendai.


\appendix

\section{Fermion propagator in a constant magnetic field} 

\label{sec:PTM}

We describe here how to obtain the dressed fermion propagator in a constant 
magnetic field shown in Fig.~\ref{fig:DressedProp}. As mentioned in the text, we employ Schwinger's 
proper-time method \cite{Sch,DR,DG}, which yields the integral 
representation (\ref{eq:GDp}) with respect to the ``proper-time" $\hat\tau$:
$$
G(p|A_{\rm cl}) =  i\left( \slashed p - e \slashed A_{\rm cl} + m \right) 
\times \frac{1}{i}
\int_0^\infty \!\! d\hat\tau \ 
{\rm e}^{ i \hat\tau \left\{ (\slashed p -e \slashed A _{\rm cl})^2 -
 (m^2-i\varepsilon ) \right\} } 
\, ,
$$
where $A_{\rm cl}^\mu$ is the Fourier transformation of 
$A_{\rm cl}^\mu(x)$ and a small imaginary value $-i\varepsilon$, 
originated from the prescription of the boundary condition, 
makes the integral convergent. 
An advantage 
of this expression is that the terms containing 
the gauge field appear in positive powers, while in the original 
definition (\ref{eq:Gp2}) the gauge field appears in the denominator. 
Below, we analyze the following integral part:
\begin{eqnarray}
\Delta(p|A_{\rm cl}) \equiv \frac{1}{i}
\int_0^\infty \!\! d\hat\tau \ {\rm e}^{ i \hat\tau \left\{ (\slashed p -e \slashed A_{\rm cl} )^2 - (m^2-i\varepsilon ) \right\} }
\ \ ,
\label{eq:Dp} 
\end{eqnarray}
which, equivalently to Eq.~(\ref{eq:Green}), satisfies the following equation:
\begin{eqnarray}
\left\{ \left(\slashed p - e \slashed A_{\rm cl} \right)^2 - m ^2 \right\} \Delta(p|A_{\rm cl}) = {\bf 1} 
\, .
\label{eq:eqDp} 
\end{eqnarray}
Note that the operator contains Dirac matrices, and thus both the equation
and $\Delta(p|A_{\rm cl})$ are understood as matrices with respect to spinor 
indices.

It is convenient to perform calculation in the Fock-Schwinger gauge
for the external field, $x_\mu A_{\rm cl}^\mu(x) = 0$, which allows one to 
represent the gauge field in terms of the field-strength tensor as 
$ A^\mu(x) =  - \frac{1}{2} F^{\mu\nu}  x_\nu$ when the field strength 
is constant. 
An elementary calculation yields the expression for Eq.~(\ref{eq:eqDp})  
in terms of the field strength: 
\begin{eqnarray}
 \left[ p^2 - \kappa^2 
+ \frac{e^2}{4} \frac{\partial}{\partial p^\mu}F^{\mu\nu}F_{\nu \sigma} \frac{\partial}{\partial p_\sigma} 
\right ] \Delta(p|A_{\rm cl}) = {\bf 1}
\ \ ,
\label{eq:DELp}
\end{eqnarray}
where we have introduced $\kappa$ defined by 
$$
\kappa^2 = m^2 - i \epsilon + \frac{e}{2} F^{\mu\nu} \sigma_{\mu\nu}\, , 
$$
with the antisymmetric tensor, 
$\sigma^{\mu\nu} = \frac{i}{2} [ \gamma^\mu , \gamma^\nu ]$. 

By solving Eq.~(\ref{eq:DELp}), one can derive another expression 
alternative to Eq.~(\ref{eq:Dp}), 
which is favorably written with respect to the field-strength tensor. 
Namely, we find 
\begin{eqnarray}
\Delta(p|A_{\rm cl}) = \frac{1}{i} \int_0^\infty \!\! d\hat\tau \ 
{\rm e}^{-i\kappa^2 \hat\tau + i p_{\mu} X^{\mu \nu} (\hat\tau) p_\nu 
+ Y(\hat\tau) }
\, ,
\label{eq:ansatz}
\end{eqnarray}
where we have introduced two functions $X^{\mu\nu}(\tau)$ 
and $Y(\tau)$ defined by
\begin{eqnarray}
X^{\mu \nu}(\hat\tau) &\equiv & 
\left[ (eF)^{-1} \tanh(eF\hat\tau) \right] ^{ \mu \nu}
= \hat\tau \ \eta_\parallel^{\mu\nu}   +  
\frac{ \tan(eB \hat\tau)}{e B} \ \eta_\perp^{\mu\nu} \, ,
\label{eq:X}
\\
Y(\hat\tau) &\equiv & 
- \frac{1}{2} {\rm Tr} \left[ \ln \left\{ \cosh(eF\hat\tau) \right\} \right] 
= - \ln \left| \cos(eB\hat\tau) \right|
\, .
\label{eq:Y}
\end{eqnarray}
In this expression, we have suppressed contracted tensor indices, 
and written inverse field-strength tensor as $(F^{-1}) ^{\mu \nu }$. 
As shown above, two functions $X^{\mu\nu}(\hat\tau)$ and $Y(\hat\tau)$ 
are especially simplified when there is only a magnetic field:
the field-strength tensor is then given by $F^{21} = - F^{12} = B$ 
with all the other elements vanishing. 
We find that $X^{\mu\nu}$ shows different behavior in the 
longitudinal ($\parallel$) and transverse ($\perp$) planes 
with respect to the external field, which are, respectively, 
specified by the metric tensors
$\eta^{\mu\nu}_\parallel = {\rm diag}(1,0,0,-1)$ and 
$\eta^{\mu\nu}_\perp = {\rm diag}(0,-1,-1,0)$.

Substituting the above expressions (\ref{eq:ansatz}) -- (\ref{eq:Y}) into Eqs.~(\ref{eq:GDp}), 
one obtains the following integral representation of the 
dressed fermion propagator:
\begin{eqnarray}
G(p|A_{\rm cl}) =  \int _0^\infty \!\! d\hat\tau 
\Big[  \slashed p -  e \gamma^\mu F_{\mu \nu} X^{\nu \sigma}(\hat\tau) 
p_\sigma + m \Big] 
{\rm e}^{-i\kappa^2 \hat\tau + i p_\mu X^{\mu \nu}(\hat\tau) p_\nu 
+ Y(\hat\tau) }\, .
\label{eq:electron}
\end{eqnarray} 
Recall that $\kappa^2$ contains the gamma matrices, 
and thus we have to take into account the order among them 
when we carry out the trace of spinor indices.  
Although integration over the proper-time $\hat\tau$ has not been performed, 
this compact form allows us to efficiently simplify awkward expressions in 
applying to resummed diagram calculations which include the 
vacuum polarization tensor shown in Fig.~\ref{fig:photon_vp}. 
Owing to the Fock-Schwinger gauge, 
a configuration of the external field is manifestly specified in terms of 
the field-strength.




\section{Renormalization of the vacuum polarization tensor}

\label{subsec:renorm}

Below we first discuss renormalization of the polarization tensor 
in the ordinary vacuum, and then consider the case with the magnetic field. 

Our framework for nonzero magnetic field background, of course, contains 
the case of the ordinary vacuum as a limit of vanishing magnetic field. 
This is explicitly 
checked in the expressions (\ref{eq:vp0}) -- (\ref{eq:Gam0}) which have not been
expanded with respect to the Landau levels. 
Indeed, in a vanishing magnetic field limit $\Br = 0$, the scalar functions 
$\chi_i  \ (i=0,1,2)$  in Eqs.~(\ref{eq:vp0}) -- (\ref{eq:Gam0}) 
reduce to the results in the ordinary vacuum. This is easily seen 
as follows.
Since $\Gamma_1$ and $\Gamma_2$ vanish in the vanishing field limit 
$\Br \rightarrow 0$ with $\tau$ and $\beta$ being fixed,\footnote{
One should take this limit after pulling back the proper-time $\tau$ 
to the original dimensionful form, $\tau \rightarrow | eB| \tau$,  
because the dimensionless form varies as the field-strength varies.}
the scalar functions $\chi_1$ and $\chi_2 $ also vanish in this 
limit: 
$$
\chi_1 = \chi_2 = 0
\, .
$$
On the other hand, $\chi_0$ reduces to the following form, 
\begin{eqnarray}
\chiv (r^2) = 
\frac{\alpha}{4\pi} \int_{-1}^1 \!\! d\beta  \int_\delta^\infty \!\! d\tau \ 
\frac{ 1- \beta^2 }{ \tau } 
{\rm e}^{ - i  \phi_0 \tau }
\ \ ,
\label{eq:chi0_vac}
\end{eqnarray}
where we have defined a phase 
$\phi_0 (r^2) =  1 - (1-\beta^2) r^2 $ with a normalized squared 
momentum $ r^2 = q^2/(4m^2) $. 
We have introduced a cut-off $\delta \, (0<\delta \ll 1)$ at the lower boundary of the integral with respect to $\tau$, 
because, without regularization $\delta \rightarrow 0$, 
there arises a logarithmic divergence in this integral. 
In terms of $\chiv$, 
the vacuum polarization tensor in the ordinary vacuum $\Pi_0^{\mu\nu} (q^2)$, 
obtained as the vanishing field limit, is given by 
\begin{eqnarray}
\Pi_0^{\mu\nu} (q) = - \chiv (r^2) P_0^{\mu\nu}
\ \ .
\end{eqnarray}

When the squared momentum $r^2$ is less than the threshold\footnote{
Expression in the other range is obtained by an analytic continuation 
afterward, if necessary.}, $0 < r^2 < 1$, 
the phase $\phi_0$ becomes positive definite, 
and then the integral with respect to $\tau$ in Eq.~(\ref{eq:chi0_vac}) 
is calculated as 
\begin{eqnarray}
\chiv &=&
\frac{\alpha}{4\pi} \int_{-1}^1 \!\! d\beta (1-\beta^2) \Gamma (0, \phi_0 \delta)
\nonumber
\\
& = &
-
\frac{\alpha}{4\pi} \int_{-1}^1 \!\! d\beta (1-\beta^2) 
\left\{
\gamma_{\rm E} + \ln \phi_0 + \ln \delta + {\cal O}(\delta^1)
\right\}
\label{eq:chi0_vac2}
\end{eqnarray}
where, on the first line, we have changed the contour so that we can use the 
`incomplete gamma function',  
$\Gamma(s,t) = \int_t^\infty x^{s-1} {\rm e}^{-x} dx$,  
and, on the second line, we have expanded it with respect to the small 
cut-off scale $\delta$. The constant $\gamma_{\rm E}$ denotes the Euler 
number. 
Note that a divergence in the limit $\delta \rightarrow 0$ is isolated 
from the other finite terms. 

Adopting the conventional on-shell renormalization condition, 
we define a finite vacuum polarization tensor $\hat \Pi_0 ^{\mu\nu} (q) $ in the ordinary vacuum as, 
\begin{eqnarray}
\hat \Pi_0 ^{\mu\nu} (q) \equiv \Pi_0  ^{\mu\nu} (q) - \Pi_0^{\mu\nu} (0) 
\, .
\end{eqnarray}
From the above on-shell condition and the expression (\ref{eq:chi0_vac2}) for 
$\chiv$, it is natural to introduce a finite scalar function $\chivr (r^2)$ 
by 
\begin{eqnarray}
\chivr (r^2) &\equiv& \chiv(r^2) - \chiv(0)
\label{eq:chi_0vac_r}  \\
&=& \frac{\alpha}{4\pi} \int_{-1}^1 \!\! d\beta (1-\beta^2) 
\ln \left\{ \frac{ 4m^2 }{4m^2 - (1-\beta^2) q^2 } \right\}\, .
\label{eq:chi0_fin0}
\end{eqnarray}
Namely, the finite vacuum polarization tensor is given by $\chivr(r^2)$:
\begin{eqnarray}
\hat \Pi_0^{\mu\nu} (q) = - \chivr(r^2)  P_0^{\mu\nu} \, ,
\end{eqnarray}
which agrees with the conventional result (see, e.g, Eq.~(7.91) in 
Ref.~\cite{PS}). 
An integral in Eq.~(\ref{eq:chi0_fin0}) with respect to $\beta$ is carried out analytically, 
and then we find an analytic form, which allows for an analytic continuation to the regimes $r^2<0$ and $1<r^2$, as 
\begin{eqnarray}
\hat \chi_0^{\rm vac} (r^2) = 
\frac{\alpha}{3\pi} \left[
\frac{1}{3} + \left( 2 + \frac{1}{r^2} \right) 
\left\{ \left( \frac{1 }{r^2} - 1 \right)^{\frac{1}{2}} \,  {\rm Arccot} \left( \frac{1}{r^2} -1 \right)^{\frac{1}{2}} - 1 \right\}
\right]
\ \ .
\end{eqnarray}

Now let us turn to the case with magnetic fields. 
As mentioned below Eqs.~(\ref{eq:vp0}) -- (\ref{eq:Gam0}), 
the vacuum polarization tensor in the presence of the external magnetic field 
contains exactly the same logarithmic divergence as the one shown in 
Eq.~(\ref{eq:chi0_vac2}). 
Therefore, we remove the logarithmic divergence by imposing a similar on-shell renormalization condition, 
\begin{eqnarray}
\hat \Pi_{\rm ex}^{\mu\nu} (q_\parallel, q_\perp; \Br) \equiv
\Pi_{\rm ex}^{\mu\nu} (q_\parallel, q_\perp; \Br) - \Pi_0^{\mu\nu} (q^2 = 0) 
\, ,
\end{eqnarray}
so that the finite vacuum polarization tensor 
$\hat \Pi_{\rm ex}^{\mu\nu} (q_\parallel, q_\perp; \Br) $ vanishes 
at simultaneous limits of $\qp=0$, $\qt=0$ and $\Br=0$. 
In terms of the scalar functions $\chi_i$, 
the renormalization condition is written as 
\begin{eqnarray}
\hat \Pi_{\rm ex}^{\mu\nu} (q_\parallel, q_\perp; \Br) =
- \left\{ \chi_0 (\rp,\rt; \Br) - \chi_0^{\rm vac} (0) \right\}  P_0^{\mu\nu}
- \sum_{i=1,2} \chi_i P_i^{\mu\nu}
\ \ .
\end{eqnarray}
To proceed further, we have to regularize the divergent scalar functions 
$\chi_0  $ and $\chi_0^{\rm vac} $.

We show regularization methods which are involved in 
our numerical \cite{HI2} and analytic calculations. 
First, in case of numerical calculation, 
we regularize $\chi_0  $ and $\chi_0^{\rm vac} $ 
by inserting common cut-off scales into these functions. 
Then, the scalar function $\chi_0  $ in the presence of the external 
magnetic field is regularized in the same manner as in 
Eq.~(\ref{eq:chi0_vac}), 
\begin{eqnarray}
\chi_0 (\rp,\rt; \Br)  =
\frac{\alpha}{4\pi} \int_{-1}^1 \!\! d\beta  \int_\delta^\infty \!\! d\tau \  
\frac{ \Gamma_0 }{ \sin\tau } {\rm e}^{ - i \phi \tau }
\ \ ,
\end{eqnarray}
where we have put all the phase appearing in Eq.~(\ref{eq:vp0}) together as 
$\phi (\rp,\rt) = \phi_\parallel(\rp) + ( u \cos(\beta \tau) - \eta \cos \tau ) / \tau$. 
However, unlike the case in the ordinary vacuum (\ref{eq:chi0_vac2}), 
a complex structure of $\chi_0  $ prevents us from isolating the divergence 
by an analytic calculation. 
We therefore numerically evaluate a convergent integral, 
\begin{eqnarray}
J_0 (\rp,\rt; \Br) = 
\frac{\alpha}{4\pi} \int_{-1}^1 \!\! d\beta  \int_\delta^\infty \!\! d\tau \  
\left[ \frac{ \Gamma_0(-i\tau) }{ \sinh\tau } {\rm e}^{ - \phi(\rp,\rt) \tau } 
- \frac{ 1 - \beta^2 }{ \tau} {\rm e}^{- \phi_0(0) \tau } \right]
\ \ ,
\end{eqnarray}
where we have rotated the integral contour downward in the complex $\tau$-plane 
assuming a momentum regime $\rp < (1+\sqrt{1+2\Br})^2/4$ 
so that the integrand vanishes on the large arc. 
We then obtain a finite part of the scalar function as, 
\begin{eqnarray}
\hat \chi_0 (\rp,\rt;\Br) &\equiv&
\chi_0  (\rp,\rt;\Br) - \chi_0^{\rm vac} (0)
\label{eq:chi_0_r}
\\
&=& J_0 (\rp,\rt; \Br) + \R_0 (\delta )
\label{eq:R0}
\ \ .
\end{eqnarray}
We have put a remainder as
\begin{eqnarray}
\R_0 (\delta )
= \frac{\alpha}{4\pi} \int_{-1}^1 \!\! d\beta  \int_0^\delta \!\! d\tau \  
\left[ \frac{ \Gamma_0 (-i\tau)}{ \sinh\tau } {\rm e}^{ -  \phi(\rp,\rt) \tau } 
- \frac{ 1 - \beta^2 }{ \tau} {\rm e}^{- \phi_0(0) \tau } \right]
\end{eqnarray}
which gives a cut-off dependence on the right-hand side of Eq.~(\ref{eq:R0}). 
Note that, if we could remove the cut-off scale 
by taking a vanishing limit $\delta \rightarrow 0$ at the end of calculation, 
the renormalized scalar function $\hat \chi_0 (\rp,\rt;\Br)$ 
should not depend on the cut-off scale 
as long as we insert common cut-off scales into $\chi_0  $ and $\chi_0^{\rm vac} $. 
However, we cannot take this limit in numerical calculations, 
and thus it should be estimated how much 
the renormalized scalar function $\hat \chi_0$ depends on the cut-off scale $\delta$. 
For a small $\tau$, the integrand is composed of the following functions, 
\begin{eqnarray}
\frac{\Gamma_0 (-i\tau)}{\sinh \tau} 
&\sim&
\frac{1-\beta^2}{\tau} + \frac{1}{6} (1-\beta^2)^2 \tau + {\cal O}(\tau^3)
\\
\phi(\rp,\rt) 
&\sim&
\phi_0(r^2) + {\cal O}(\tau^3)
\ \ ,
\end{eqnarray}
and thus the remainder is evaluated as, 
\begin{eqnarray}
\R_0 (\delta) &\sim& 
\frac{\alpha}{4\pi} \int_{-1}^1 \!\! d\beta  \int_0^\delta \!\! d\tau \ 
\left[
- (1-\beta^2) \left\{ \phi_0(r^2)  - \phi_0(0) \right\} + {\cal O}(\tau)
\right]
\label{eq:integ_tau}
\\
&=&
 \frac{4\alpha}{15\pi} r^2  \delta + {\cal O}(\delta^2)
\ \ .
\end{eqnarray}
The above estimate shows that the remainder $\R_0 (\delta)$ is linearly suppressed as 
we take a small cut-off scale $\delta$ in numerical calculation.

There is an improved method to suppress the cut-off scale dependence, 
which we have adopted in our numerical calculation. 
Note that the term independent of $\tau$ in Eq.~(\ref{eq:integ_tau}) 
vanishes, if two phases $\phi_0$ are evaluated at the same momentum. 
We rearrange Eq.~(\ref{eq:chi_0_r}) to eliminate this constant term 
by using Eq.~(\ref{eq:chi_0vac_r}) as
\footnote{
An equivalent expression to Eq.~(\ref{eq:chi_0_r1}) has been 
shown in Ref.~\cite{MS76}, and might be adopted to numerical 
computation in Ref.~\cite{KY}. 
However, it has not been clear how it improves the naive subtraction 
(\ref{eq:chi_0_r}). 
}
\begin{eqnarray}
\hat \chi_0 (\rp,\rt;\Br) = 
\left\{ \chi_0  (\rp,\rt;\Br) - \chi_0^{\rm vac} (r^2) \right\}
+ \hat \chi_0^{\rm vac} (r^2) 
\label{eq:chi_0_r1}
\ \ .
\end{eqnarray}
The renormalized scalar function is then obtained as, 
\begin{eqnarray}
\hat \chi_0 (\rp,\rt;\Br) = 
J (\rp,\rt; \Br) + \hat \chi_0^{\rm vac} (r^2) + \R (\delta )
\ \ ,
\end{eqnarray}
with a convergent integral $J (\rp,\rt; \Br)$ and a remainder $\R(\delta)$ given by 
\begin{eqnarray}
J (\rp,\rt; \Br) &=&
\frac{\alpha}{4\pi} \int_{-1}^1 \!\! d\beta  \int_\delta^\infty \!\! d\tau \  
\left[ \frac{ \Gamma_0 (-i\tau)}{ \sinh\tau } {\rm e}^{ - \phi(\rp,\rt) \tau } 
- \frac{ 1 - \beta^2 }{ \tau} {\rm e}^{-  \phi_0(r^2) \tau } \right]\, ,
\\
\R (\delta ) &=&
\frac{\alpha}{4\pi} \int_{-1}^1 \!\! d\beta  \int_0^\delta \!\! d\tau \  
\left[ \frac{ \Gamma_0(-i\tau) }{ \sinh\tau } {\rm e}^{ -  \phi(\rp,\rt) \tau } 
- \frac{ 1 - \beta^2 }{ \tau} {\rm e}^{- \phi_0(r^2) \tau } \right]
\, .
\end{eqnarray}
We already have an analytic expression for the renormalized scalar function 
$ \hat \chi_0^{\rm vac} (r^2) $ in the ordinary vacuum. 
In this method, 
the remainder $\R(\delta)$ is evaluated as, 
\begin{eqnarray}
\R (\delta) &\sim& 
\frac{\alpha}{4\pi} \int_{-1}^1 \!\! d\beta  \int_0^\delta \!\! d\tau \  
\left[
\left\{ \frac{1}{2} (1-\beta^2) \phi_0(r^2) - \frac{1}{6} (1-\beta^2)^2 \right\} \tau  
+ {\cal O} (\tau^2)
\right]
\\
&=&
\frac{\alpha}{180 \pi} (11 - 12 r^2) \delta^2 + {\cal O}(\delta^3)
\, .
\end{eqnarray}
By using this method, 
the cut-off dependence is improved 
from a linear dependence $\R_0(\delta) \sim {\cal O}(\delta^1)$ to 
quadratic one $\R(\delta) \sim {\cal O}(\delta^2)$.

Next, we proceed to the case of our analytic calculation. 
In the expansion method resulting in the double infinite sum with respect to the Landau levels, 
we do not have to insert a cut-off scale to perform term-by-term integrals, 
because these contributions, 
from the discretized levels in the transverse momentum, does not diverge individually. 
However, by taking all-order summation, we show 
that the infinite sum leads to a logarithmic divergence 
which should be identified with the logarithmic divergence we have seen above.

We examine the double infinite series at vanishing momenta $\rp = 0$ and $\rt= 0$, 
because the ultraviolet divergence itself does not depend on external momentum scales. 
At these momenta, 
the series coefficient $C_\ell^n$ and its multiplicative form in $\chi_0$ have simple expressions, 
\begin{eqnarray}
C_\ell^n ( 0 ) &=& 
\left\{
\begin{array}{ll}
 1 & (n =0 )
\\
0 & (n \geq 1 )
\end{array}
\right.
\\
\frac{n}{\eta} C_\ell^n ( 0 ) &=& 
\left\{
\begin{array}{ll}
\ell + 1 & (n =1 )
\\
0 & (n \geq 2) 
\ \ ,
\end{array}
\right.
\end{eqnarray}
where we have used $L_\ell^0 (0) =1$ and $ L_\ell^1 (0) = \ell + 1$. 
Note that the expression of $\chi_0$ should be understood 
that any term being proportional to $ n/\eta \times C_\ell^n ( 0 )$ 
at $n=0$ is absent, 
because such term never exists in the expansion (\ref{eq:pwd3}). 
Then, summation of the coefficients in $\chi_2$ (see Eqs.~(\ref{eq:D1}) and (\ref{eq:D2})) 
are also easily performed as 
\begin{eqnarray}
D_\ell^{0 (1)} (0)  &=& -8 \sum_{\lambda=0}^{\ell-1}
(\ell - \lambda)  \left\{ - C_\ell^0(0) \right\}
= 4 \ell(\ell+1)
\ \ ,
\\
D_\ell^{1 (1)} (0)  &=& -8 \sum_{\lambda=0}^{\ell-1}
(\ell - \lambda)   C_\ell^0(0)
= -4 {\ell}(\ell+1)
\ \ ,
\\
D_\ell^{2 (1)} (0)  &=& -8 \sum_{\lambda=0}^{\ell-2}
(\ell - \lambda-1)  \left\{ - C_\ell^0(0) \right\}
= 4 {\ell} (\ell-1)
\ \ .
\end{eqnarray}
Therefore, the infinite series with respect to $n$ terminates up 
to the first two terms. 
At vanishing momenta and for $n = 0,1$, 
the functions $F_\ell^n(0)$, $G_\ell^n(0)$ and $H_\ell^n(0)$ have 
the following forms:
\begin{eqnarray}
F_\ell^0 (0) &=&  \frac{ 2 }{ 1 + 2 \ell \Br }\, ,
\\
F_\ell^1 (0) &=& - \frac{1}{\Br} \Xi_\ell^1\, ,
\\
G_\ell^1 (0) &=&  - \frac{1}{\Br^2} 
\left[ 
\left\{ 1 + ( 2 \ell + 1) \Br \right\} \Xi_\ell^1 + 2 \Br
\right]\, ,
\\
H_\ell^0 (0) &=& \frac{ 2 }{3} \cdot \frac{1}{ 1 + 2 \ell \Br }
\, .
\end{eqnarray}
By using these explicit expressions, 
we show that the scalar coefficient $\chi_0$ and differences $\chi_1-\chi_0$ and $\chi_2-\chi_0$ 
contain the same logarithmic divergence. 
Introducing $\tilde \chi_i = \chi_i - ( 1 - \delta_{i 0} ) \chi_0$, 
we have 
\begin{eqnarray}
\tilde \chi_i (\rp=\rt=0;\Br) = 
\frac{\alpha \Br}{4\pi} 
\left\{ u_i (\Br)  + \sum_{\ell=1}^{\infty} w_i (\ell; \Br) \right\}
\ \ ,
\end{eqnarray}
where we separated an infinite series with respect to $\ell$ 
so that it begins from $\ell=1$. 
Explicit expressions of terms are given by 
\begin{eqnarray}
u_i (\Br) = 
\left\{
\begin{array}{ll}
2 \left( F_0^1 - G_0^1 \right) & (i=0)
\\
F_0^0 + F_{1}^0 - ( H_0^0 + H_{1}^0 ) & (i=1)
\\
0 & (i=2)
\end{array}
\right.
 \ \,
\end{eqnarray}
and 
\begin{eqnarray}
w_i (\ell ; \Br) = 
\left\{
\begin{array}{ll}
2 \left\{ F_\ell^1 - (2\ell+1) G_\ell^1 \right\}  & (i=0)
\\
F_\ell^0 + F_{\ell+1}^0 - ( H_\ell^0 + H_{\ell+1}^0 ) & (i=1)
\\
4 \ell (\ell+1) \left\{ 
( F_\ell^0   - 2 F_\ell^1 ) + F_{\ell+1}^0  
\right\} & (i=2)
\end{array}
\right.
\ \ .
\end{eqnarray}
Here, we suppressed the vanishing arguments of $F_\ell^n(0)$, $G_\ell^n(0)$ and $H_\ell^n(0)$. 
To see a behavior of the summation taken up to a large $\ell$, 
we insert a cut-off at $\ell = L $ assuming an infinite limit $L \rightarrow \infty$. 
The infinite sum is then evaluated by being replaced by an integral as, 
\begin{eqnarray}
\tilde \chi_i (\rp=\rt=0; \Br)
&= &
\frac{\alpha \Br}{4\pi} 
\left\{ u_i (\Br)  + 
 \lim_{L\rightarrow \infty} L \sum_{\ell=1}^{L} \frac{1}{L} 
w_i ( \ell ; \Br) \right\}
\nonumber \\
&=&
\frac{\alpha \Br}{4\pi} 
\left\{ u_i (\Br)  + 
\lim_{L\rightarrow \infty} L \int_0^1 \!\! dx \, w_i ( Lx ; \Br) \right\}
\ \ .
\end{eqnarray}
The above integrals are carried out analytically, 
and then expanded for a large value of $L$ as, 
\begin{eqnarray}
\tilde \chi_i (\rp=\rt=0; \Br)
=
\frac{\alpha}{3\pi} \ln(2 \Br L)
+ \sum_{ k=0 }^{\infty}  c_k(\Br) L^{-k}
\label{eq:logdiv}
\ \ ,
\end{eqnarray}
where series coefficients are denoted by $c_k(\Br)$. 
The above expansion series does not have any positive power of $L$, 
and thus the leading term for large $L$ (with finite $\Br$) 
is given by a divergent logarithm. 
Note also that a coefficient of the divergent term is independent of $\Br$, 
implying the same divergence as in the ordinary vacuum. 
While the scalar functions $\chi_1$ and $\chi_2$ do not contain any divergence 
owing to a cancellation between the logarithmic terms in the functions $\tilde \chi_i $, 
the other one $\chi_0$ contains the logarithmic divergence found in Eq.~(\ref{eq:logdiv}). 
These results are consistent with the observation below 
Eq.~(\ref{eq:vp0}) -- (\ref{eq:Gam0}). 
Therefore, we might be allowed to identify this divergent logarithm, appearing from the infinite sum, 
with the one in Eq.~(\ref{eq:chi0_vac2}) appearing in the limit $\delta \rightarrow 0$. 
Comparing them up to the divergent logarithmic term as 
\begin{eqnarray}
\frac{\alpha}{3\pi} \ln( \delta^{-1} ) = \frac{\alpha}{3\pi} \ln ( 2 \Br L )
\ \ ,
\end{eqnarray}
we then find a relation between the cut-off scales, 
\begin{eqnarray}
\delta^{-1} = 2 \Br L
\label{eq:Cscales}
\ \ .
\end{eqnarray}
According to this relation (\ref{eq:Cscales}), 
the ultraviolet logarithmic divergence in the presence of a magnetic field 
is subtracted by the same logarithmic divergence in the ordinary vacuum. 
Namely, the divergent contribution in a sum up to 
$\ell = L \gg 1 $ might require to be subtracted by an integral 
(\ref{eq:chi0_vac}) evaluated with a cut-off scale $\delta = (2\Br L)^{-1}$, 
and a finite part in our prescription is defined as a remainder of this 
subtraction.

\section{Dispersion relations from modified photon propagator}  
\label{sec:Pprop}

Here we demonstrate that one can obtain Eq.~(\ref{eq:eigen}) 
(that immediately yields the dispersion relations 
(\ref{eq:eps1bperp}) and (\ref{eq:eps1bparallel})) 
from pole positions of a modified photon propagator 
after taking into account the interactions with the external magnetic field. 

\begin{figure}
	\begin{center}
         \includegraphics[scale=0.35, clip]{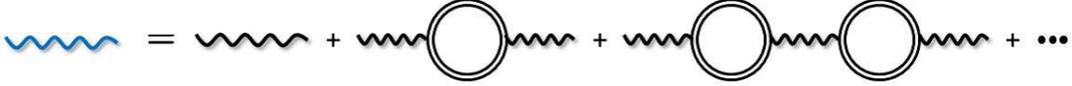}
\caption{A modified photon propagator
}
	\end{center}
\label{fig:Dphoton}
\end{figure}

First of all, it should be noticed that insertion of the 
self-energy $\Pi_{\rm ex}^{\mu\nu}$ into a photon propagator 
corresponds to resumming the dressed-fermion one-loop diagrams as 
shown in Fig.~7. 
There, the dressed-fermion one-loops are 
linked by the free photon propagators $D_0^{\mu\nu} (q^2)$ given by
\begin{eqnarray}
iD_0^{\mu\nu}(q) \equiv \frac{1}{q^2} 
\left[ 
\eta^{\mu\nu} - ( 1 - \xi_g ) \frac{ q^\mu q^\nu }{ q^2 }
\right] 
\, .
\label{eq:D0}
\end{eqnarray}
A full photon propagator in this approximation can be expressed as
\begin{eqnarray}
D^{\mu\nu} (q) = \sum_{n=0}^\infty D_n^{\mu\nu}(q) 
\label{eq:Dsum}
\, ,
\end{eqnarray}
where the $n$-th contribution corresponds to the diagram having $n$ rings, 
and $n=0$ is the free propagator $D_0^{\mu\nu}$.
To compute the $n$-th order contribution $D_n^{\mu\nu}(q^2) $, 
we use the following identities among the projection operators 
$P_i^{\mu\nu}$ : 
\begin{eqnarray}
&&
   P_0^{\mu\rho}  P_{i\rho}^{\ \ \nu}  
=  P_i^{\mu\rho}  P_{0\rho}^{\ \ \nu} 
= q^2  P_i^{\mu\nu} \ \ (i=0,1,2)\, ,
\nonumber
\\
&&
P_1^{\mu\rho}  P_{1\rho}^{\ \ \nu}  = q_\parallel^2 P_1^{\mu\nu}
\ , \ \ 
P_2^{\mu\rho}  P_{2\rho}^{\ \ \nu} =  q_\perp^2 P_2^{\mu\nu} 
\ , \ \ 
   P_1^{\mu\rho}  P_{2\rho}^{\ \ \nu} 
=  P_2^{\mu\rho}  P_{1\rho}^{\ \ \nu} = 0
\ \ .
\end{eqnarray}
Since, as shown by the above relations, the products of two projection operators always result in zero or 
one of the projection operators, 
we are able to expand the $n$-th order diagram 
into the same form as the self-energy with respect to the projection operators: 
\begin{eqnarray}
D_n^{ \mu\nu}(q) 
&= &
D_0^{\mu\rho_1}  ( i \Pex _{\rho_1 \sigma_1} D_0^{\sigma_1\rho_2} )
\cdots
( i \Pex _{\rho_n \sigma_n}  D_0^{\sigma_n\nu} )\nonumber
\\
&\equiv &
\zeta_0^{(n)}  P_0^{\mu\nu} + \zeta_1^{(n)}  P_1^{\mu\nu} + \zeta_2^{(n)}  P_2^{\mu\nu}
\label{eq:Dn}
\ \ . 
\end{eqnarray} 
For $n \geq 1$, 
we can drop those terms proportional to $q^{\mu} q^{\nu}$ 
in the free propagator (\ref{eq:D0}), because they vanish when 
contracted with any of the projection operators in the self-energy 
as $ q_\mu P_i^{\mu\nu} = 0 \ (i=0,1,2)$. 
The coefficient scalar functions are then calculated as 
\begin{eqnarray}
\zeta_0^{(n)} 
&=& 
\left( \frac{-i}{q^2}  \right) 
\times
\left( - \frac{\chi_0}{q^2} \right)^n  (q^2)^{n -1}\, ,
\label{eq:zeta1}
\\
\zeta_1^{(n)} 
&=& 
\left( \frac{-i}{q^2}  \right) 
\times
\sum_{r=0}^{n-1} {}_n\mathrm{C}_r \left( - \frac{\chi_0}{q^2} \right)^{r} \left( - \frac{\chi_1}{q^2} \right)^{n-r}
(q^2)^{r} \ (q_\parallel^2)^{n-1-r} 
\nonumber
\\
&=&
\left( \frac{-i}{q^2}  \right) 
\times \frac{ (-1)^n } {q_\parallel^2}
\left\{ \left( \chi_0 + \frac{q_\parallel^2}{q^2} \chi_1 \right)^n - ( \chi_0 )^n \right\} \, ,
\label{eq:zeta2}
\\
\zeta_2^{(n)} 
&=& 
\zeta_1^{(n)}
\left( 1 \rightarrow 2 \ , \ \ \parallel \rightarrow \perp \right)
\, ,
\label{eq:zeta3}
\end{eqnarray}
where the last line is meant for replacements of 
subscripts in $\zeta_1^{(n)}$ as indicated.
To carry out the summations in $\zeta_1^{(n)}$ and $\zeta_2^{(n)}$, 
we used a formula for the binomial expansion. 
Substituting Eq.~(\ref{eq:Dn}) with the coefficients 
(\ref{eq:zeta1}) -- (\ref{eq:zeta3}) 
into Eq.~(\ref{eq:Dsum}),  
we obtain the resummed photon propagator as 
\begin{eqnarray}
&iD^{\mu\nu} (q) &\nonumber\\
&=&\!\!\!\!
\frac{1}{ (1+\chi_0) q^2 } \cdot \frac{ P_0^{\mu\nu} }{ q^2 } 
+
\frac{ \sigma_\parallel (q^2,q_\parallel^2) }{ (1 + \chi_0 )  q^2  +  \chi_1 q^2_\parallel  }  
\cdot \frac{ P_1^{\mu\nu} }{ q_\parallel^2 }
+ 
\frac{ \sigma_\perp(q^2,q_\perp^2) }{  (1 + \chi_0 )  q^2  +  \chi_2 q^2_\perp } 
\cdot \frac{ P_2^{\mu\nu} }{ q_\perp^2 }
+ 
\frac{ \xi_g }{q^2}  \cdot \frac{ q^\mu q^\nu }{ q^2 } 
\nonumber
\\
&=&\!\!\!\!
\frac{  \pi_{(0)}^\mu \pi_{(0)}^\nu }{ (1+\chi_0) q^2 }
+
\frac{ \pi_{(1)}^\mu \pi_{(1)}^\nu }{ (1+\chi_0) q^2 + \chi_2 q^2_\perp } 
+
\frac{ \pi_{(2)}^\mu \pi_{(2)}^\nu }{ (1+\chi_0) q^2 + \chi_1 q^2_\parallel }
+
\frac{ \xi_g }{q^2} \pi_{(3)}^\mu \pi_{(3)}^\nu\, .
\label{eq:D}
\end{eqnarray}
We have introduced 
$ \sigma_\parallel(q^2,q_\parallel^2) \equiv - q_\parallel^2 \chi_1 / \{ q^2 (1+\chi_0) \}  $
and 
$ \sigma_\perp(q^2,q_\perp^2) \equiv - q_\perp^2 \chi_2 / \{ q^2 (1+\chi_0) \}  $, 
and utilized  relations shown in Eq.~(\ref{eq:proj}) to get 
a `diagonalized' form on the second line. 
Contracting the propagator (\ref{eq:D}) with eigenvectors as, 
$\pi_{(i)}^\mu D_{\mu\nu} \pi_{(i)}^\nu \ (i=0,1,2,3)$, 
we find that the pole position of each mode exactly agrees with 
what has been obtained in Eq.~(\ref{eq:eigen}) from  
diagonalization of the modified Maxwell equation.


\section{Physical modes of a photon in a magnetic field}

\label{sec:EM}

In this Appendix, we show that, among four polarization modes 
in Eq.~(\ref{eq:eigen}), only two modes, having nontrivial dispersions, are 
physical. We consider electric and magnetic fields of the dynamical 
photon field (\ref{eq:Aprop}) whose polarizations are specified by the eigenvectors 
$\pi_{(\lambda)}^\mu$ $(\lambda=0,\cdots,3)$ defined in 
Eq.~(\ref{pol_vectors}). Explicitly, the Fourier modes of 
the electric field ${\cal E}^i\equiv -F^{0i}$ and magnetic field 
${\cal B}^i\equiv \frac12 \epsilon^{ijk}F_{jk}$ of a propagating photon 
$a^\mu$ in Eq.~(\ref{eq:Aprop}) are, respectively, given by
\begin{eqnarray}
{\cal E}_{(\lambda)}^i(q) &=& i N \left( \omega \pi_{(\lambda)}^i 
- q^i \pi_{(\lambda)}^0 \right) 
\, {\rm e}^{-i(\omega t - \bq \cdot \bx) } 
\, ,
\label{eq:e}
\\
{\cal B}_{(\lambda)}^i(q) &=& i N 
\left( \epsilon^{ijk} q^j 
\pi_{(\lambda)}^k \right) 
\, {\rm e}^{-i(\omega t - \bq \cdot \bx) } 
\, .
\label{eq:b}
\end{eqnarray}


Let us first consider the polarization modes $\lambda = 0 \ 
{\rm and}\ 3$ (see Eq.~(\ref{eq:vec2})), both of which give the 
 ordinary dispersion relation of the massless type $\omega^2=|\bq^2|$.
Substituting the definitions of $\pi_{(0)}^\mu$ and $\pi_{(3)}^\mu$ 
into the above, one immediately finds
\begin{eqnarray}
\begin{array}{l}
\vec{\cal E}_{(0)} \propto \left(
\begin{array}{c}
q_x\\ q_y\\ 0
\end{array}
\right)\, q^2 = \vec{0}\, ,
\qquad 
\vec{\cal B}_{(0)} \propto  \left(
\begin{array}{c}
q_y\\ -q_x\\ 0
\end{array}
\right)\, q^2= \vec{0}\, ,
\end{array}
\ \ \ \ {\rm for}\ \lambda=0 \, ,\label{unphysical0}
\end{eqnarray}
and 
\begin{eqnarray}
\begin{array}{l}
\vec{\cal E}_{(3)} = \vec{0}\, ,\qquad 
\vec{\cal B}_{(3)} = \vec{0} \, , \ \ \ {\rm for}\ \lambda=3\, ,
\end{array} \label{unphysical3}
\end{eqnarray}
where the dispersion relation $q^2=0$  has been used
for the $\lambda=0$ mode. 
Namely, the two modes $\lambda=0, 3$ do not induce 
excitation of the electromagnetic fields, which implies that they are 
unphysical. In fact, these modes are related to redundant gauge degrees of 
freedom.

In contrast, the remaining two modes $\lambda=1,2$ 
provide nonvanishing electromagnetic fields. First of all, one can easily 
verify (without using the dispersion relations) that the electric field 
and the magnetic field are orthogonal to each other for each mode: 
$\vec{\cal E}_{(\lambda)} \cdot \vec{\cal B}_{(\lambda)} = 0 \ 
(\lambda=1,2)$. Next, let us employ the same coordinates as those used 
in deriving Eqs.~(\ref{eq:eps1bperp}) and (\ref{eq:eps1bparallel}). 
Then the electromagnetic field of the $\lambda=1$ mode (with momentum $q^\mu$) 
reads
\begin{eqnarray}
\begin{array}{l}
\vec{\cal E}_{(1)} (t,\bx)
= \omega N \ \left(
\begin{array}{c}
0\\  1\\  0
\end{array}
\right) \ \Psi_\perp (t,\bx)\, ,\quad
\vec{\cal B}_{(1)} (t,\bx)
= \omega N \ \left(
\begin{array}{c}
\cos \theta\\ 
0\\ 
\sin \theta
\end{array}
\right) \  \Psi_\perp (t,\bx)
\ \ ,
\end{array}
\label{eq:eb1}
\end{eqnarray}
where the common factor $\Psi_\perp (t,\bx)$ is defined by 
\begin{equation}
\Psi_\perp (t,\bx) \equiv 
{\exp}\big\{-i\omega (t - n_{\perp}^{\rm real}\, \hat{\bq} \cdot \bx )\big\} 
\, 
{\exp}\big\{-\omega n_{\perp}^{\rm imag}\, \hat{\bq} \cdot \bx \big\}\, .
\end{equation}
Note that the factor includes real and imaginary parts of the refractive
index $n_{\perp}$, which respectively represent oscillation and 
decay behavior of the electromagnetic field. Thus, they give 
a phase velocity and damping coefficient, respectively. 
A unit vector $\hat \bq$ denotes a three-dimensional vector directed 
in the photon momentum $\bq$. 
As in a photon in the ordinary vacuum, 
these fields are transverse to the photon momentum: 
$\vec{q} \cdot \vec{\cal E}_{(1)} = \vec{q} \cdot \vec{\cal B} _{(1)} = 0$. 
The electric field induced in this mode is oriented to 
the transverse direction with respect to the external magnetic field, 
and thus does not have parallel component.

Similarly, the electromagnetic field of the last remaining 
mode $\lambda=2$ is given by
\begin{eqnarray}
\hspace*{-3mm}
\vec{\cal E}_{(2)} = 
\omega N 
\sqrt{ \frac{\epsilon_\parallel (\theta)}
       {\epsilon_\parallel  (\frac{\pi}{2})} 
     } 
\left(
\begin{array}{c}
 \epsilon_\parallel \! \left( \frac{\pi}{2} \right) \cos \theta \\
 0 \\
 \sin \theta 
\end{array}
\right)  \Psi_\parallel (t, \bx) \, , \quad 
\vec{\cal B}_{(2)} = 
\omega N \sqrt{\epsilon_\parallel \! \left( \frac{\pi}{2} \right) } 
\left( 
\begin{array}{c}
0\\ -1\\ 0
\end{array}
\right)  \Psi_\parallel (t, \bx)
\,  .
\label{eq:eb2}
\end{eqnarray}
where $\epsilon _\parallel (\theta) $ denotes 
the dielectric constant when the photon momentum is directed to 
the zenith angle 
$\theta$.
While the magnetic field is transverse to the photon momentum 
$\vec{q} \cdot \vec{\cal B}_{(2)} = 0$, we notice that the electric field 
in general has an overlap with the photon momentum as, 
$\vec{q} \cdot \vec{\cal E}_{(2)} = 
\omega N \sqrt{ \frac{\epsilon_\parallel \! (\theta)}
{\epsilon_\parallel \! (\frac{\pi}{2})} } 
\left\{  \epsilon_\parallel \! \left( \frac{\pi}{2} \right)-1 \right\} \sin \theta \cos \theta$, 
depending on the propagation angle $\theta$ with respect to the external magnetic field.

The space-time dependence of the field is again represented by 
a common factor $\Psi_\parallel (t,\bx)$, and it describes oscillation and 
decay behavior of the field. However, it contains the refractive index 
$n_\parallel$ (instead of $n_\perp$). 
Namely, the factor is explicitly written as
\begin{equation}
\Psi_\parallel (t,\bx) \equiv 
\exp \big\{ - i \omega ( t - n_\parallel^{\rm real}\, \hat{\bq} \cdot \bx ) \big\} 
\exp \big\{ - \omega n_\parallel^{\rm imag}\, \hat{\bq} \cdot \bx \big\}.
\end{equation} 
Recall that the refractive index $n_\parallel$ is in general different from 
$n_\perp$ in $\Psi_\perp(t,\bx)$. Therefore, the two physical modes 
$\lambda=1,2$ have distinct phase velocities and decay rates 
(into a fermion-antifermion pair). 
The electric field shown in Eq.~(\ref{eq:eb2}) contains a component 
oscillating along the external field 
in which direction a charged particle has continuous spectrum. 
This is in contrast to the electric field of the other physical mode. 
Since the external magnetic field provides anisotropic 
spectra of fermions in the Dirac sea, oscillating electric 
fields, $\vec{\cal E}_{(1)}$ and $\vec{\cal E}_{(2)}$, 
of an incident photon 
induce anisotropic polarizations.


\section{Partial wave decomposition} 

\label{sec:pwd}

We expand exponentiated trigonometric functions in Eqs.~(\ref{eq:vp0}) -- 
(\ref{eq:Gam0}) by using the Jacobi-Anger expansion that is 
well-known as the partial-wave decomposition \cite{BK}:
\begin{eqnarray}
\exp \left\{ - i u \cos(\bt) \right \} 
&=& \sum_{n=0}^{\infty} (2-\delta_{n0}) (-i)^n J_n(u) \cos( n\bt) 
\nonumber
\\
&=& \sum_{n=0}^{\infty} (2-\delta_{n0}) I_n(-iu) \cos( n\bt) 
\nonumber
\\
&\rightarrow& 
\sum_{n=0}^{\infty} (2-\delta_{n0}) I_n(-iu) \ {\rm e}^{in\bt} \, ,
\\
\cos(\bt) \exp \left\{ - i u \cos(\bt) \right \} 
&=& i \frac{\partial\ }{\partial u} \exp \left\{ - i u \cos(\bt) \right \} 
\nonumber
\\
&=& \sum_{n=0}^{\infty} \frac{1}{2}(2-\delta_{n0}) 
\left\{ I_{n+1}(-iu) + I_{n-1}(-iu) \right\} \cos( n\bt) 
\nonumber
\\
&\rightarrow&  \sum_{n=0}^{\infty} \frac{1}{2} (2-\delta_{n0}) \left\{ I_{n+1}(-iu) + I_{n-1}(-iu) \right\} \ {\rm e}^{in\bt} \, ,
\\
\sin(\bt) \exp \left\{ - i u \cos(\bt) \right \} 
&=& \frac{1}{iu\beta} \frac{\partial\ }{\partial \tau} \exp \left\{ - i u \cos(\bt) \right \}
\nonumber
\\
&=& \frac{i}{u} \sum_{n=0}^{\infty} n (2-\delta_{n0}) I_n(-iu) \sin( n\bt) 
\nonumber
\\
&\rightarrow&  \frac{1}{u} \sum_{n=0}^{\infty} n (2-\delta_{n0}) I_n(-iu) 
\ {\rm e}^{in\bt}
\label{eq:pwd3}
\, ,
\end{eqnarray}
where $J_n(z)$ and $I_n(z)$ are the Bessel and modified Bessel functions 
of the first kind related as, $I_n(z) = (-i)^n J_n(iz)$. 
Noting that the integrands in Eq. (\ref{eq:Gam1}) are even function 
of $\beta $, we have replaced the trigonometric functions by exponential 
factors on the right-hand sides of the arrows, i.e.,  
$\cos(n\bt) \rightarrow {\rm e}^{in\bt}$ and  
$\sin(n\bt) \rightarrow - i {\rm e}^{in\bt}$.


\section{Limiting form of the coefficient $C_\ell^n(\eta)$} 


\label{sec:formulae}

An approximate form of $C_\ell^n(\eta)$ for large values of 
$\ell \gg 1$ is found by the use of a limiting form of the associated Laguerre 
polynomial and Stirling's formula for the factorial $\ell !$. 
When $\ell$ is large enough with fixed $n$ and $\eta$, 
a limiting form of the associated Laguerre polynomial $L_\ell^n(\eta)$ 
is expressed as (see (13.5.14) in Ref.~\cite{Abr}), 
\begin{eqnarray}
L_\ell^n(\eta) &\sim& 
\frac{1}{  \sqrt{\pi \ell}  } 
\left( \frac{\ell}{\eta} \right) ^{\frac{n}{2}+\frac{1}{4} }
{\rm e}^{\frac{\eta}{2} } \, 
\cos \Psi_\ell^n (\eta) 
\label{eq:Llimit}
\, ,
\\
\Psi _\ell^n (\eta) &=& 
2 \sqrt{ \eta \left( \ell + \frac{n+1}{2} \right) } - \frac{\pi}{2} \left( n+\frac{1}{2} \right) 
\label{eq:Llimit2}
\, .
\end{eqnarray}
Also, we use 
Stirling's formula for large $\ell$ 
(see (6.1.38) in Ref.~\cite{Abr}), 
\begin{eqnarray} 
\ell! \sim \sqrt{ 2 \pi } \,  {\rm e}^{-\ell}\, \ell^{ \, \ell + \frac{1}{2} } 
\label{eq:Stirling}
\ \ .
\end{eqnarray}
Then, the ratio of factorials in $C_\ell^n(\eta)$ for large $\ell \gg 1$ 
reads, 
\begin{eqnarray}
\frac{ \ell ! }{ (\ell+n)! }
&\sim&
{\rm e}^n \ell^{-n} \left( \frac{\ell}{\ell+n} \right) ^{ \ell+n+\frac{1}{2} } 
\nonumber
\\
&\sim&
{\rm e}^{- \frac{n+1}{2\ell} } \ell^{-n}
\label{eq:Rlimit}
\ \ ,
\end{eqnarray}
where we have used 
$ \left\{ \ell/(\ell+n) \right\}^{\ell+n+1/2} 
= {\rm e}^{ -(\ell+n+1/2) \ln(1+n/\ell)} 
\sim {\rm e}^{ - n \left\{ 1 + (n+1)/(2\ell ) \right\} } $ 
to obtain the final expression. 
By using Eqs.~(\ref{eq:Llimit}), (\ref{eq:Llimit2}) and (\ref{eq:Rlimit}), 
we find a limiting form of $C_\ell^n(\eta)$ as, 
\begin{eqnarray}
C_\ell^n(\eta) &\sim& \frac{1}{\pi \sqrt{ \eta\ell \  } } 
\, {\rm e}^{ -\frac{n+1}{2\ell} } \cos^2 \Psi_\ell^n (\eta) 
\ \qquad ( \ell \gg 1)
\label{eq:Clim}
\, .
\end{eqnarray}
The amplitude of $C_\ell^n(\eta)$ is bounded by a prefactor of the squared 
cosine 
which decays by an inverse-square-root as $\ell $ becomes 
large (see Fig.~\ref{fig:C}).

\section{Imaginary part of the function $I_{\ell \Delta}^n (r_\parallel^2)$} 

\label{sec:atan}

In this Appendix, we carefully investigate analytic properties of the 
function $I_{\ell \Delta}^n (r_\parallel^2)
=I_{\ell +}^n (r_\parallel^2)-I_{\ell -}^n (r_\parallel^2)$ (see Eq.~(\ref{eq:Ipm})) 
which is the only one possible source of the imaginary contribution 
and thus determines the essential behavior of the imaginary part of the vacuum polarization tensor.
In the following, we consider the function $I_{\ell \Delta}^n (r_\parallel^2)$
with the indices $\ell$ and $n$ being fixed.

As mentioned in the text, the property of $I_{\ell \Delta}^n (r_\parallel^2)$
is specified by the discriminant ${\cal D}$ defined by 
Eq.~(\ref{discriminant}). Using the solutions $\rp=s_\pm^\idx$ to the equation 
${\cal D}=0$ (see Eq.~(\ref{eq:s_pm})), we consider three kinematical regions 
(for notational simplicity, we suppress the indices $\ell$ and $n$ in $s_\pm^\idx$):
\begin{eqnarray}
({\rm I})\qquad\qquad \rp<s_- &\qquad&({\cal D}<0)\nonumber\\
({\rm II})\quad\ s_-<\rp<s_+ &\qquad& ({\cal D}>0) \\
({\rm III})\quad\ s_+<\rp \hspace*{9mm}&\qquad& ({\cal D}<0)\, .\nonumber
\end{eqnarray}
In region (II) where ${\cal D}>0$, $I_{\ell \Delta}^n (r_\parallel^2)$ is 
obviously a real-valued function of $r_\parallel^2$, and thus 
there arises no imaginary part. In regions (I) and (III), the argument 
of the cotangent in $I_{\ell \pm}^n (r_\parallel^2)$ becomes a pure imaginary number. 
The expression of $I_{\ell \pm}^n (r_\parallel^2)$ in these regions is available via analytic
continuation of the function in region (I). 
With an infinitesimal imaginary displacement $\rp + i\epsilon$, 
analytic continuation into the regions (I) and (III)
leads to the following form valid in both regions: 
\begin{eqnarray}
I_{\ell \pm}^n (r_\parallel^2 + i \epsilon) 
&=& 
\frac{-2i}{ \sqrt{b^2-4ac} }
\arctan \left\{ -i ( w_\pm + i \epsilon \kappa_\pm ) \right\}
\nonumber
\\
&=&
\frac{1}{ \sqrt{b^2-4ac} } 
\left\{ \ln ( 1 - w_\pm -  i \epsilon \kappa_\pm ) - \ln( 1 + w_\pm +  i \epsilon \kappa_\pm ) \right\}
\label{eq:atan_app}
\end{eqnarray}
where we have introduced two real-valued functions,  
$w_\pm (r_\parallel^2) = (b \pm 2a)/\sqrt{b^2-4ac}$ and 
$\kappa_\pm (r_\parallel^2) = b \mp 2 a $. 
The latter functions $\kappa_\pm$ are leading-order coefficients 
in an expansion of $w_\pm (r_\parallel^2 + i \epsilon)$ 
with respect to the infinitesimal displacement. 
Owing to the expression in terms of the logarithm, we notice that the arctangent has an imaginary part 
when the absolute values of $w_\pm$ are larger than one, $|w_\pm|>1$, 
and that its sign can be either positive or negative depending on the sign of $\kappa_\pm$.

First, we show in what region $I_{\ell \pm}^n (r_\parallel^2)$ has an 
imaginary part. 
Comparing the denominator and numerator of $w_\pm$ as 
$(b \pm 2a )^2 - \sqrt{b^2-4ac}^2 = 4 \rp \left\{ 1 + ( 2\ell + n \mp n ) \Br \right\}$, 
we obtain inequalities $|w_\pm|>1$ and $|w_\pm| \leq 1$, 
when the longitudinal momentum squared is positive $\rp>0$ and negative $\rp \leq 0$, respectively. 
Therefore, $I_{\ell \pm}^n (r_\parallel^2)$ has an imaginary part 
only in the positive regime, $\rp > 0$.

Next, we should notice that the difference $I_{\ell \Delta}^n (\rp) = I_{\ell +}^n (\rp) - I_{\ell -}^n (\rp) $ has an imaginary part 
only when the functions $I_{\ell \pm}^n(r_\parallel^2)$ have imaginary 
parts with opposite signs: otherwise, they cancel each other. 
The difference $I_{\ell \Delta}^n (\rp)$ retains an imaginary part 
when the arguments of the arctangent, $w_\pm + i \kappa_\pm$, are on the opposite sides of the branch cut. 
Thus, we shall examine the signs of $w_\pm$ and $\kappa_\pm$ in more detail, 
which simply follow from those of $b+2a$ and $b-2a$ according to their definition below Eq.~(\ref{eq:atan_app}).

In the following, the signs of the arguments are found in the regions, 
$0<\rp<s_-$ and $s_+<\rp \ (0 \leq s_- < s_+)$, separately. 
First, we find an obvious inequality, $b-2a=-( n \Br + 2 \rp)<0$, as long as $\rp > 0$. 
Thus, the signs of $\sgn(w_-)$ and $ \sgn (\kappa_+) $ immediately 
follows in the both regions as 
\begin{eqnarray}
\sgn(w_-) < 0 \ \ {\rm and}\ \ \sgn (\kappa_+) < 0 
\ \ \ (0 < \rp)
\label{eq:sign1}
\ \ .
\end{eqnarray}
On the other hand, the signs of the other functions depend on the momentum regime. 
To see this, we show two relevant inequalities valid in the each regime, 
\begin{eqnarray}
&&
b + 2a < - n \Br + 2s_-  
< ( 1 + 2 \ell \Br ) -  \sqrt{1+2\ell\Br}^2 
= 0 
\ \ \ (0 < \rp \leq s_-)
\label{eq:ineq1}
\ \ ,
\\
&&
b + 2a > - n \Br + 2s_+ 
> - n \Br + \frac{1}{2} \sqrt{ 1 + 2(\ell+n) \Br }^2
> 0
\ \ \ (s_+ \leq \rp)
\ \ .
\label{eq:ineq2}
\end{eqnarray}
Led by these inequalities, we obtain the signs of the remaining two functions depending on the regime: 
\begin{eqnarray}
&&
\sgn(w_+) < 0 \ \ {\rm and } \ \ \sgn (\kappa_-) < 0 
\ \ \ (0 < \rp < s_-)
\label{eq:sign21}
\ \ ,
\\
&&
\sgn(w_+) > 0 \ \ {\rm  and}\ \ \sgn (\kappa_-) > 0  
\ \ \ (s_+ < \rp)
\label{eq:sign22}
\ \ .
\end{eqnarray}

Remind that the branch cut of the arctangent is located on the imaginary axis. 
From Eqs.~(\ref{eq:sign1}) and (\ref{eq:sign21}), 
we find that the imaginary parts of $I_{\ell \pm}^n (r_\parallel^2)$ have the same signs in the regime $0<\rp<s_-$, 
and that the function $I_{\ell \Delta}^n (r_\parallel^2)$ does not have any imaginary part in this regime. 
On the other hand, from Eqs.~(\ref{eq:sign1}) and (\ref{eq:sign22}), 
we find that the imaginary parts of $I_{\ell \pm}^n (r_\parallel^2)$ have the opposite signs in the regime $s_+<\rp$, 
and that the function $I_{\ell \Delta}^n (r_\parallel^2)$ has an imaginary part in this regime. 
Therefore, we conclude that 
the difference $I_{\ell \Delta}^n(r_\parallel^2)$ retains an imaginary part only in the regime $s_+<\rp$, namely, in region (III). 
A piecewise representation of $I_{\ell \Delta}^n(r_\parallel^2)$ is shown in Eq.~(\ref{eq:I}). 
It has the same form for the real part in regions (I) and (III), 
but has an imaginary part only in region (III). 

The imaginary part in the regime $s_+^\idx < \rp$ is related to 
the analytic property of the vacuum polarization tensor in the complex $\rp$-plane. 
It reflects that there are an infinite number of branch cuts initiating at 
the thresholds of photon decay, $\rp = s_+^\idx$, specified by pairs of Landau-level indices $\ell$ and $\ell+n$, 
where a created fermion and antifermion pair has vanishing longitudinal momenta along the external magnetic field. 
Because the longitudinal momenta of a created fermion pair is continuous, 
each branch cut runs continuously along the real axis into the asymptotic regime, $\rp \rightarrow \infty$.

\section{Limiting behavior of $I_{\ell \Delta}^n (\rp)$ at the boundaries}

\label{sec:limit}

In this Appendix, we discuss the limiting behavior of 
$I_{\ell \Delta}^n (\rp)$ in Eq.~(\ref{eq:I}) 
at the boundaries, $\rp = s_\pm$ 
(we suppress the indices $\ell$ and $n$ for simplicity). 
We put a square root of the discriminant as 
$ \delta = \sqrt{|b^2-4ac|} $ 
which vanishes at both of the boundaries. 
For infinitesimal $\delta$, the logarithm and arctangent in 
$I_{\ell \Delta}^n (\rp)$ behave as 
$ \log \left| \frac{a-c-\delta}{a-c+\delta} \right|
 = \frac{2}{c-a} \delta + {\cal O}(\delta^3)$ 
and $ \arctan \{( b \pm 2 a ) /\delta \} = \frac{\pi}{2} \cdot \sgn(b\pm2a) - \frac{1}{b \pm 2a} \delta + {\cal O}(\delta^3)$, respectively. 
Thus, the limiting behavior of $I_{\ell \Delta}^n (\rp)$ is represented as 
\begin{eqnarray}
\lim_{\rp \rightarrow s_- , s_+} I_{\ell \Delta}^n (\rp)
= \left\{
\begin{array}{l}
\frac{2}{c-a} + {\cal O}( \delta^2 )
\ \ \qquad \qquad \quad (\rp \rightarrow s_- - 0)
\\
\left\{ \sgn (b+2a) - \sgn (b-2a)  \right\} \pi \delta^{-1}
\\
\hspace{0.5cm}
+ \frac{2}{c-a} + {\cal O} ( \delta^2 )
\ \ \qquad  \quad (\rp \rightarrow s_- + 0 \ {\rm and} \ s_+ - 0 )
\\
2 \pi i \delta^{-1} + \frac{2}{c-a} + {\cal O} ( \delta^2 )
\ \ \quad (\rp \rightarrow s_+ + 0 )
\end{array}
\right.
\end{eqnarray}
where we have used $b^2 \rightarrow 4 ac \ (\delta \rightarrow 0)$ 
to obtain the zeroth-order term in the middle regime. 
Note that a negative-power term in the middle regime 
exists depending on a relative sign between the factors $b \pm 2a$. 
In Appendix \ref{sec:atan}, we have already examined 
those signs just above Eq.~(\ref{eq:sign1}) and in Eqs. (\ref{eq:ineq1}) and (\ref{eq:ineq2}). 
They have the same signs as $b+2a<0$ and $b-2a<0$ 
in the vicinity of the lower boundary $\rp = s_- \pm 0$, 
while the opposite signs as $b+2a>0$ and $b-2a<0$ 
in the vicinity of the higher boundary $\rp = s_+ \pm 0$.
\footnote{
Remind that the inequality in Eq.~(\ref{eq:ineq1}) does not contain an equality, 
and has been shown including at the boundary $\rp = s_-$. 
Noting that the factor $b+2a $ is regular with respect to $\rp$ at the boundary, 
we find that it has the same sign on the both sides of the boundary 
as far as its infinitesimal vicinity is concerned. 
The same is true for the other inequality in Eq.~(\ref{eq:ineq2}) 
in the infinitesimal vicinity of the boundary $\rp = s_+$. 
} 
Therefore, $I_{\ell \Delta}^n (\rp)$ is finite in the limit $\rp \rightarrow s_- + 0$, 
and, on the other hand, $I_{\ell \Delta}^n (\rp)$ is divergent in the limit $\rp \rightarrow s_+ - 0$.

The limiting behaviors of $I_{\ell \Delta}^n (\rp)$ are summarized as follows. 
At $\rp = s_- $, it is continuous, 
and thus has a limiting value $I_{\ell \Delta}^n (\rp \rightarrow s_-) = \left. 2/(c-a) \right|_{\rp = s_-}$. 
On the other hand, 
$I_{\ell \Delta}^n (\rp)$ diverges to the positive infinity 
in the limit $\rp \rightarrow s_+ - 0 $. 
The real part of $I_{\ell \Delta}^n (\rp)$ has a limiting value 
$I_{\ell \Delta}^n (\rp \rightarrow s_+ + 0) = \left. 2/(c-a) \right|_{\rp = s_+}$ 
in the limit $\rp \rightarrow s_+ + 0 $, 
while the imaginary part diverges to the positive infinity in this limit.




\end{document}